\documentclass[pdflatex,sn-mathphys-num]{sn-jnl}
\usepackage{fullpage}
\usepackage{setspace}
\usepackage{placeins}
\usepackage{graphicx}
\usepackage{multirow}
\usepackage{amsmath,amssymb,amsfonts}
\usepackage{amsthm}
\usepackage{mathrsfs}
\usepackage[title]{appendix}
\usepackage{xcolor}
\usepackage{textcomp}
\usepackage{manyfoot}
\usepackage{booktabs}
\usepackage{algorithm}
\usepackage{algorithmicx}
\usepackage{algpseudocode}
\usepackage{listings}
\usepackage{subfig}

\begin{document}

\title[Chopping and distilling variational autoencoders for real-time anomaly detection in high energy physics]{Chopping and distilling variational autoencoders for real-time anomaly detection in high energy physics}

\author[1]{\fnm{Max} \sur{Cohen}}
\equalcont{\setstretch{1}These authors contributed equally to this work.}

\author[2]{\fnm{Rajat} \sur{Gupta}}
\equalcont{\setstretch{1}These authors contributed equally to this work.}

\author[1]{\fnm{Sterre} \sur{Hoogendoorn}}
\equalcont{\setstretch{1}These authors contributed equally to this work.}

\author[2]{\fnm{Sergey} \sur{Scoville}}

\author[2]{\fnm{Tae Min} \sur{Hong}}\email{tmhong@pitt.edu}

\author[1]{\fnm{Dylan Sheldon} \sur{Rankin}}\email{dsrankin@sas.upenn.edu}

\affil[1]{\setstretch{1.0}\orgdiv{Department of Physics \& Astronomy}, \orgname{University of Pennsylvania}, Philadelphia, PA 19104}

\affil[2]{\setstretch{1.0}\orgdiv{Department of Physics \& Astronomy}, \orgname{University of Pittsburgh},
Pittsburgh, PA 15260}

\vspace{-2em}%
\abstract{Anomaly detection (AD)  has recently emerged as an exciting alternative to conventional search strategies in high energy physics using artificial intelligence (AI) and machine learning (ML). The integration of these techniques into trigger systems is even more recent, but represents a crucial step in expanding the coverage of LHC triggers. In this paper, we explore the direct comparison, as well as combination, of two compression techniques for variational autoencoder (VAE) AD trigger algorithms: utilizing only latent-space derived variables and therefore requiring only half of the VAE that we call ``chopping'' and applying knowledge distillation (KD) to distill the VAE into a student architecture that we call ``distillation.'' We demonstrate the feasibility of deploying such techniques on an FPGA within the resource and latency constraints of an LHC trigger environment and further find that a combination of the two leads to the smallest models that maintain, and in some cases, improve, performance with respect to the original VAE architecture.
}


\maketitle

\setcounter{tocdepth}{2}
\vspace{-2em}%
\tableofcontents

\section{Introduction}
\label{sec:introduction}

While the Standard Model (SM) of particle physics is successful, it is known to be incomplete. The lack of a particle description for dark matter, or a theoretical explanation for the observed matter-antimatter asymmetry or the hierarchy problem, suggests the presence of additional, undiscovered particles. Searches for new particles in high energy physics (HEP) have historically first selected particular models of new physics and then optimized selections to target these models. This strategy enabled the discovery of the Higgs boson in 2012 at the Large Hadron Collider (LHC) at CERN~\cite{ATLAS-HIGG-2012,CMS-HIGG-2012}, and has helped produce hundreds of searches for new particles. However, the lack of another discovery from these searches suggests the need to revisit this strategy. The use of anomaly detection (AD)\footnote{Throughout the paper, some abbreviations and unusual terms are noted in the margin.}\marginpar{AD} methods has emerged as a potential way to remove any bias introduced by the physicist's selection of a particular model to target in searches. Such methods of searching for new physics at the LHC have begun to produce results in recent years, and offer an important complementary approach to traditional search methods. AD methods continue to be developed and refined in the context of HEP, and a particularly novel direction has been the introduction of AD into HEP triggers.

Trigger systems in HEP are responsible for quickly reducing the rate of data that needs to be collected by implementing fast filtering that rejects uninteresting data before writing it to disks. At the LHC, and in particular in ATLAS and CMS, these triggers rely initially on field-programmable gate arrays (FPGA) to reduce event rates from 40 MHz to 1-0.1 MHz, all within $\mathcal{O}(1)\,\mu s$. Traditional algorithms rely on fast, simple, reconstruction of detector signals and cut-based selection methods to identify interesting objects and reject events. While these algorithms are often designed without specific new physics models in mind, they are unconsciously biased towards simpler events for which it is easy to understand the backgrounds and signals. For example, ATLAS and CMS implement one and two electron trigger algorithms aimed at identifying events that produced one and two electrons, but do not implement four electron trigger algorithms, or two electron, two muon trigger algorithms. While it is possible to design traditional triggers to target specific exotic cases, the use of AD in triggers offers a compelling way to target them more generally. In recent years, both ATLAS and CMS have deployed multiple AD-based trigger algorithms, based either on the deployment only of subsets of trained models or on the distillation of large trained models into much smaller ones. In spite of this, there is no consensus on the advantages and disadvantages of these different approaches, or the ramifications in terms of their performance. We present a range of different techniques aimed at reducing the size of anomaly detection algorithms, with the ultimate goal of making them implementable at ultra-low latencies for HEP triggering applications of FPGA.

The paper is organized as follows. Section~\ref{sec:related} describes the related work in AD and HEP. Section~\ref{sec:method} explains the methods we employ along with the datasets we consider. The results from our studies are presented in Section~\ref{sec:result}. We end by summarizing our findings and commenting on future research questions in Section~\ref{sec:conclusion}.

\section{Related Work}
\label{sec:related}

\subsection{Anomaly Detection in High Energy Physics}

Along with other artificial intelligence (AI) and machine learning (ML) techniques, AD has gained popularity in HEP in recent years. Given that labeled data are not necessary for many AD techniques, it has become a common choice for those looking to perform model-agnostic searches for physics beyond the Standard Model (BSM).  A dedicated community challenge, the LHC Olympics~\cite{Kasieczka:2021xcg}, has been instrumental in benchmarking and advancing these approaches across a wide range of methods and signal topologies. Broadly, AD methods in HEP can be categorized by the assumptions they make about the signal: weakly supervised methods such as CWoLa~\cite{Metodiev_2017}, ANODE~\cite{Nachman:2020lpy}, and CATHODE~\cite{Hallin:2021wme} assume the signal is localized in a resonant feature, gaining sensitivity at the cost of this prior assumption, while fully unsupervised methods such as autoencoders make no such assumption, trading sensitivity for broader applicability. A third category is offered by semi-supervised methods, which incorporate partial prior knowledge about expected signal topology without committing to a specific model. We elaborate on the three categories below.

\paragraph{Weakly supervised resonant methods}
Weakly supervised resonant AD methods search for localized overdensities in a signal region by comparing data to a data-driven background model. CWoLa, ANODE, and CATHODE are foundational early examples of this approach, scanning a resonant feature (e.g., invariant mass) and using a conditional density estimator trained on sidebands to model the background in a signal region, identifying overdensities by comparing the data to the interpolated background model. Subsequent works have extended this framework by exploring different strategies for constructing the background template against which the signal region data is compared. SALAD~\cite{Andreassen:2020nkr} uses simulation as a reference, reweighting it to match the data in the sidebands before interpolating into the signal region. CURTAINs~\cite{Raine:2022hht} takes a purely data-driven approach, using a normalizing flow to learn how features change across the mass spectrum and applying this to transport sideband events directly into the signal region. FETA~\cite{Golling:2022nkl} combines these ideas, using a flow to morph simulated events into a realistic background template in the signal region, benefiting from simulation as a physics-informed prior while avoiding the limitations of reweighting when simulation and data differ significantly.

These weakly supervised resonant methods have been deployed in actual experimental searches on LHC data. The ATLAS collaboration published the first weakly supervised dijet resonance search using the CWoLa method~\cite{ATLAS:2020iwa}, and subsequently an extended search employing the SALAD reweighting and CURTAINs morphing strategies  with jet substructure features over the full Run~2 dataset~\cite{ATLAS:2025obc}. The  CMS collaboration similarly performed a model-agnostic dijet search that included  weakly supervised components based on CWoLa Hunting, CATHODE, and Tag N' Train algorithms~\cite{CMS:2024nsz}. No significant deviations from the Standard Model  were observed in any of these searches.

\paragraph{Fully unsupervised methods}
While the methods discussed above assume the signal is localized in a resonant feature such as invariant mass, a complementary class of approaches makes no such assumption. Self-supervised and fully unsupervised methods instead aim to identify anomalous events based purely on how unusual they appear relative to the bulk of the data, without any prior knowledge of where in phase space a signal might appear.

A natural and  widely-used architecture for this task is the autoencoder (AE)\marginpar{AE,VAE} and its probabilistic  extension, the variational autoencoder (VAE). The underlying idea is that a model trained exclusively on Standard Model background events will learn to reconstruct typical events with low loss, while anomalous events, which were not represented  during training, will incur a higher reconstruction error, providing an anomaly  score. This approach was pioneered in HEP by Farina et al.~\cite{Farina:2018fyg}, who demonstrated that deep autoencoders can flag anomalous jet images, and was  subsequently extended to VAEs and event-level data by Cerri et al.~\cite{Cerri:2018anq}, who proposed deploying such an algorithm in the LHC  trigger system. VAEs have since been applied to a range of tasks including  anomalous jet tagging~\cite{Cheng:2020dal} and event-level new physics searches. More recent works have studied the limitations of reconstruction-based anomaly detection: Finke et al.~\cite{Finke:2021sdf} showed that standard autoencoders can fail to tag anomalies in a model-agnostic way due to the sparsity and topology of jet images, arguing that the choice of architecture and input representation must be made carefully. Further theoretical investigation by Batson et al.~\cite{Batson:2021agz} demonstrated that topological mismatches between signal and background can lead to systematic failures for reconstruction-based scores, motivating alternatives such as the normalized autoencoder~\cite{Dillon:2022mkq}, which uses an energy-based model to avoid assigning anomalously high reconstruction probability to genuinely complex Standard Model events. Beyond AEs and VAEs, normalizing flows, diffusion models, and other generative architectures have also been explored for unsupervised AD in HEP~\cite{Kasieczka:2021xcg}.

On the experimental side, the ATLAS collaboration has published two fully unsupervised searches on LHC collision data: a search for new phenomena in two-body invariant mass distributions across multiple final states using an unsupervised autoencoder~\cite{ATLAS:2023ixc}, and a search for new resonances decaying into a Higgs boson and a generic new particle in hadronic final states, also using an autoencoder to define anomaly regions~\cite{ATLAS:2023azi}. The CMS collaboration's model-agnostic dijet search additionally included a fully unsupervised VAE-based component with mass decorrelation~\cite{CMS:2024nsz}. No significant excesses were  observed in any of these searches.

\paragraph{Semi-supervised methods}
A third category is offered by semi-supervised methods such as QUAK~\cite{Park:2020pak}, which incorporate partial prior knowledge about the expected signal topology without committing to a specific model. Rather than searching blindly or assuming a resonance, these methods embed physics-motivated inductive biases into the anomaly score, improving sensitivity to broad classes of new physics while retaining model-agnosticity. The CMS model-agnostic dijet search also deployed a semi-supervised component based on QUAK~\cite{CMS:2024nsz}, making it the only experimental result to date to compare unsupervised, weakly supervised, and semi-supervised anomaly detection methods within a single analysis. For a comprehensive survey of the broader landscape of AD methods in HEP, we refer the reader to Ref.~\cite{Belis:2023mqs}.

\subsection{Neural Networks in FPGAs} \label{subsec:NNs_in_FPGAs}
The LHC trigger systems run on FPGAs and therefore impose tight latency and resource constraints on the trigger algorithms. This follows a general trend of modern developments in ML and AI increasing the need for real-time data analysis on dedicated hardware systems such as FPGAs, ASICs, and GPUs. As such, techniques and tools are being widely developed to facilitate the deployment of ML models---in particular neural networks (NN)---onto highly constrained hardware environments. This development can largely be broken down into two categories: compression of a model to meet small resource and timing constraints and functional methods for converting Python-based models for implementation onto an FPGA.

\paragraph{Model Compression}

A common first step before FPGA implementation is to compress the neural network---that is, to reduce the number of operations needed for an inference step. One common technique is known as quantization \cite{jacob2018quantization,gholami2022survey}, where model weights and activations are converted to lower-precision representations (e.g. 32-bit to 8-bit). This can either be done after training is complete, known as post-training quantization (PTQ), or during training itself using quantization aware training (QAT). Another common technique, known as pruning \cite{han2015learning}, removes unnecessary weights or connections from a neural network. Low-rank factorization, on the other hand, decomposes weight matrices into products of lower-rank matrices thereby eliminating redundant operations. A canonical example of this approach is explored by Denton et al \cite{denton2014exploiting}. Knowledge distillation \cite{hintonDistillingKnowledgeNeural2015} can also be thought of as a compression scheme. With this technique, a smaller network---the student---is trained to approximate the output of a larger network---the teacher. We will cover knowledge distillation in greater detail in Section \ref{subsec:KD}. In practice, one often combines several approaches in order to achieve the best balance between compression and performance.

\paragraph{FPGA conversion}
From the FPGA architecture perspective, there are now many frameworks that exist to help convert Python-based ML code to code usable on an FPGA---many of which directly integrate various compression techniques. One such framework commonly used within HEP is HLS4ML \cite{Schulte:2025mai}. This is an open-source compiler that converts ML models from common deep learning frameworks such as PyTorch and Keras to HLS code that can be deployed on an FPGA. HLS4ML also provides built-in functionality for tuning resource and latency usage of models through parallelization, variable precision, implementation strategy, and integration with pruning and quantization tools. For a survey of other frameworks to convert NN algorithms to FPGA architectures see \cite{boutrosFieldProgrammableGateArray2025}.

HLS4ML is commonly used within workflows for CMS and ATLAS trigger algorithm design \cite{francescatoModelCompressionSimplification2021,Gandrakota:2024hbo,Govorkova:2021AEFPGA,gerlach2025evaluationnovelfastmachine}. Within these examples, most authors utilize some combination of compression techniques, such as pruning, quantization, and knowledge distillation,  along with the functionality provided by HLS4ML to further reduce NN size.

With these techniques and tools widely available, the process of developing and deploying a NN-based architecture on an FPGA within a constrained system like the LHC trigger system is becoming increasingly feasible. However, challenges remain in balancing architecture size and design with efficiency and accuracy. To develop such a model still requires considerable time and effort to explore and choose the best architectures, compression techniques, and implementation strategies to use.

\subsection{Boosted Decision Trees in FPGAs}

Boosted decision trees (BDT) are a long-standing machine-learning method in HEP and have been widely used in both offline analyses and fast classification tasks. Their structure also makes them appealing for low-latency trigger environments implemented on FPGAs. Inference with a BDT consists of a sequence of threshold comparisons followed by an accumulation of leaf responses. Compared with neural networks, which are dominated by matrix multiplications and therefore tend to require substantial DSP resources, BDT rely primarily on comparators, lookup logic, and simple additions. This makes them a natural candidate for FPGA deployment in systems where latency and logic utilization are the dominant constraints. At the same time, the hierarchical structure of tree traversal introduces its own challenge: ordinary decision-tree inference is intrinsically serial, since each event must move through a branching sequence from root to leaf. Considerable work has therefore gone into restructuring BDT inference so that it can be implemented more efficiently in hardware.

\paragraph{Model compression}

From the perspective of hardware deployment, BDT compression is closely tied to the structure of the ensemble itself. The principal model parameters that govern hardware cost are the number of trees, the maximum tree depth, and the numerical precision used to represent thresholds and leaf values. Reducing the number of trees lowers the total number of decision paths that must be evaluated, while limiting the depth shortens the effective inference path length and reduces the number of comparisons needed per tree. Likewise, fixed-point quantization of thresholds and leaf outputs can significantly reduce logic and memory cost while preserving most of the predictive performance of the trained model. These forms of compression are especially natural for BDT because the learned model is already expressed as a piecewise tabular structure rather than a dense matrix of weights.

In HEP applications, BDT have often been used not only as standalone classifiers, but also as compact surrogates for more complex models. In this sense, they can themselves serve as a form of knowledge distillation: instead of deploying a larger neural network directly, one may train a BDT regressor or classifier to approximate its outputs using the same input representation. This makes BDT particularly attractive in trigger environments, where one seeks a model that is both small and interpretable while retaining as much as possible of the teacher model's discriminating power. The trade-off, however, remains model-dependent. Increasing the ensemble size or depth can improve the approximation fidelity, but does so at the cost of larger firmware implementations and potentially longer effective latency. As with neural-network compression, identifying the optimal BDT configuration therefore requires balancing hardware simplicity against physics performance.

\paragraph{FPGA conversion}

A number of frameworks now exist to translate trained BDT models into firmware suitable for FPGA deployment. Within HEP, \textsc{TMVA} remains a widely used training framework for boosted trees~\cite{TMVA:2007ngy}, and its outputs are commonly used as the starting point for hardware-oriented conversion flows. One important development is the extension of \textsc{hls4ml}, originally designed for NNs, to support tree-based models~\cite{Summers:2020}. In this approach, the trained forest is translated into a high-level synthesis representation in which the branch decisions and leaf outputs are implemented directly in FPGA logic. This enables deployment of BDT within the same general toolchain that has become standard for many NN trigger studies.

Another important line of development is provided by \textsc{fwXmachina}~\cite{Hong:2021, Carlson:2022dgb, Serhiayenka:2024han}, which targets ultra-low-latency FPGA inference for tree-based models. Rather than preserving the conventional sequential tree-walk structure, \textsc{fwXmachina} restructures trained forests into a flattened, tabularized representation in which many threshold comparisons can be evaluated in parallel. This transformation is carried out through intermediate steps such as tree flattening and forest merging, after which the resulting model can be exported into firmware-oriented forms including high-level synthesis or VHDL implementations. By turning tree traversal into a largely combinational binning problem, this approach can substantially reduce cycle latency relative to naive tree inference and is therefore well suited to real-time trigger systems operating on nanosecond time scales.

Taken together, these developments make BDT an increasingly practical model class for low-latency HEP applications. They occupy a useful middle ground between the expressive power of NN-based anomaly detection models and the strict resource and timing requirements of FPGA trigger hardware. For this reason, BDT provide a natural comparison point, and in some cases a natural student architecture, in studies of compressed anomaly detection algorithms for real-time deployment.

\subsection{Knowledge Distillation}
\label{subsec:KD}

As mentioned previously in the discussion of NN, knowledge distillation (KD)\marginpar{KD} is a common model compression technique, where knowledge from a larger, well-trained \emph{teacher}\marginpar{\emph{teacher}} model is transferred into a much smaller \emph{student}\marginpar{\emph{student}} model. The term ``knowledge distillation'' was first popularized by \cite{hintonDistillingKnowledgeNeural2015}, although earlier work on model compression was done by \cite{bucilaModelCompression2006} and \cite{baDeepNetsReally2014}.  Since then, many different KD techniques have emerged with the common goal of exploring how best to define and distill different forms of knowledge that may exist in the teacher beyond simply the outputs or a target derived from them. For example, some techniques include: training students using information from both output and intermediate layers (“hints layers”) from the teacher \cite{romeroFitNetsHintsThin2015}; progressive distillation through teacher assistant networks to bridge the capacity gap between the student and the teacher \cite{mirzadehImprovedKnowledgeDistillation2020}; combining quantization with knowledge distillation \cite{polinoModelCompressionDistillation2018}; self-distillation \cite{zhangBeYourOwn2019,jangSelfDistilledSelfSupervisedRepresentation2022}; distillation of relation-based knowledge \cite{parkRelationalKnowledgeDistillation2019}; and multi-teacher distillation \cite{youLearningMultipleTeacher2017}. For a more comprehensive survey of KD techniques see \cite{gouKnowledgeDistillationSurvey2021}. 

KD is also becoming a more common compression technique within the context of the LHC. For example, \cite{Pol:2023dxw} presented a KD procedure with co-learning (training the teacher and the student simultaneously) and outlier exposure (exposing the student network to outliers/signals to increase generalization to higher scores) for training an anomaly detection algorithm. Later, CICADA, \cite{Gandrakota:2024hbo}, a student model regressed to the log transformed output of a CNN VAE, was developed for deployment as an AD trigger in Run 3 of CMS. Further exploration of CICADA was done in \cite{gerlach2025evaluationnovelfastmachine} by combining KD methods with outlier exposure and various quantization techniques. In \cite{francescatoModelCompressionSimplification2021}, quantized KD and hint layers were used as part of a multi-stage compression framework to adapt a toy ATLAS muon trigger neural network for deployment onto an FPGA. As these examples demonstrate, and, as discussed in Section \ref{subsec:NNs_in_FPGAs}, the development of techniques for model compression, such as KD, is becoming an increasingly important aspect of algorithm design within the constrained trigger environments of ATLAS and CMS.

\section{Method}
\label{sec:method}

\subsection{Compression Schemes}

\begin{figure}
    \centering
    \includegraphics[width=1.1\linewidth]{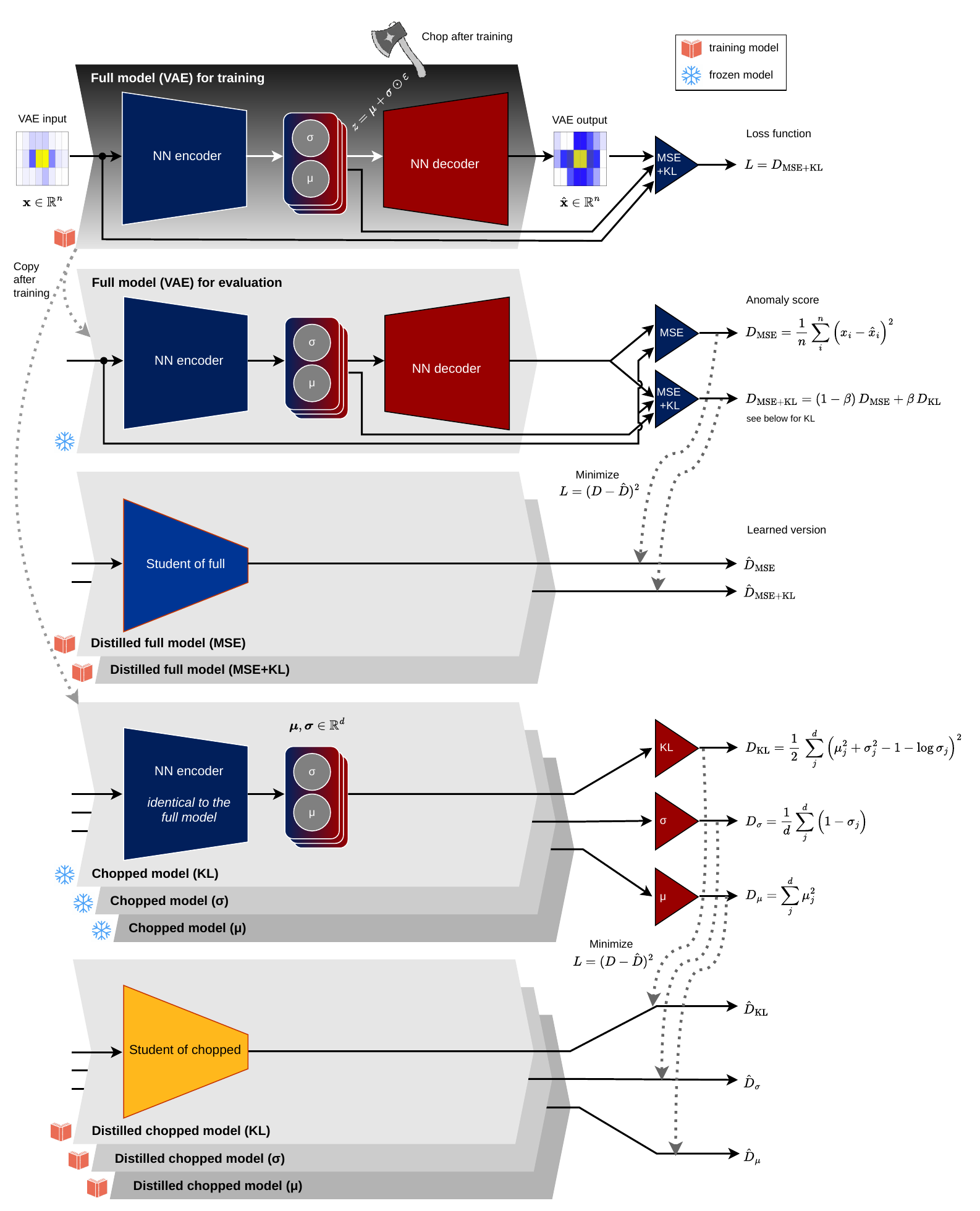}
    \caption{Schematic of the different compression schemes. Anomaly scores $D$ and the learned values $\hat{D}$ listed here are defined in Equations \ref{eq:vaeloss}--\ref{eq:mu_score}.}
    \label{fig:schemes}
\end{figure}

The full VAE model can be compressed by distillation or by chopping. The latter can be further distilled itself. A schematic of the different schemes is shown in Fig.~\ref{fig:schemes}. The figure at the top shows the setup for training the full VAE model. The training procedure is highlighted by a gradient background, which uses the mean squared error (MSE)\marginpar{MSE} between the input variables $\mathrm{x}$ and the output variables $\hat{\mathrm{x}}$, both quantities in $\mathbb{R}^n$, as the loss function to minimize. The figure in the second row is the frozen model after the training procedure, where the coefficients are fixed for evaluation. In this scenario two anomaly scores can be computed denoted as $D$. The MSE can be considered by itself as $D_\mathrm{MSE}$. It can also be combined with a Kullback-Leibler (KL)\marginpar{KL} divergence to form the score $D_\mathrm{MSE+KL}$. Each anomaly score of the frozen full model can be distilled, denoted as $\hat{D}$, as shown in the third row of figures. The figure in the fourth row is the chopped model, where the encoder is frozen. It can be configured to produce three anomaly scores, each of which can be distilled, as shown in the last row. The formulae given in the figure are described later.

\paragraph{``Chopping'' the AD in the latent space}

In many low-latency settings such as ATLAS and CMS triggers, traditional network compression techniques alone are not sufficient to meet timing and resource requirements. One way to address this, which has been gaining popularity in recent years, is performing the anomaly detection task in the latent space of the VAE rather than using metrics such as the input-output MSE. If the anomaly score uses exclusively latent-space variables, the decoder half of the VAE is not necessary during inference. Thus, this set of methods is called \emph{chopping} \marginpar{\emph{chop}} in this paper. 

In the context of low-latency trigger applications, chopping was introduced by Govorkova et al.~\cite{Govorkova:2021AEFPGA}, who emphasized that implementing the full VAE inference procedure on FPGA hardware is challenging, particularly because the reparameterization step requires random number generation. To avoid this difficulty, they proposed anomaly scores based exclusively on latent-space quantities, which also removes the need to evaluate the decoder during inference. They proposed to use either the full KL divergence as the anomaly score, or a z-score metric: 

\begin{equation}
R_z = \sum_i \frac{\mu_i^2}{\sigma_i^2}
\label{equation:rz}
\end{equation}

This use of exclusively latent-space variables for anomaly scoring differed from earlier work which often used reconstruction-based variables. For example, Cerri et al. \cite{Cerri:2018anq} use the full loss, MSE+KL, as the anomaly score, and An and Cho \cite{AnCho:2015VAEAD} propose an estimate of the reconstruction probability.

This strategy was utilized and further refined by an AD trigger algorithm at CMS known as AXOL1TL~\cite{Zipper:2023jqo, Gandrakota:2024hbo}. AXOL1TL uses $\sum_j \mu_j^2$ as the anomaly score, corresponding to a further simplification of Eq.~\eqref{equation:rz}. 

For our study, we propose to study five different anomaly scores based on the previous literature and AD trigger algorithms.
Two scores depend on the output of the autoencoder $\hat{\mathbf{x}}\in\mathbb{R}^n$:
\begin{align}
    &D_\text{MSE+KL} = (1-\beta)\,D_\text{MSE} + \beta\,D_\text{KL},~\textrm{where} 
\label{eq:vaeloss}
\end{align}
\begin{align}
    D_\text{MSE} = \frac{1}{n} \sum_{i}^n \big(x_i-\hat{x}_i\big)^2
    \label{eq:mse_score}
\end{align}
and three scores depend only on the latent space variables $\boldsymbol{\mu},\boldsymbol{\sigma}\in\mathbb{R}^d$, which we call \emph{chopped anomaly scores} since the decoder is ``chopped off'' of the autoencoder:
\begin{align}
    &D_\text{KL} = \frac{1}{2} \sum_{j}^d \big(\sigma_j^2 + \mu_j^2 - 1 - \log{\sigma_j^2}\big)  \label{eq:KL_score}\\ 
    &D_\text{$\sigma$} = \frac{1}{d} \sum_{j}^d \big(\sigma_j - 1\big) \label{eq:sigma_score}\\
    &D_\text{$\mu$} = \sum_{j}^d \mu_j^2 \label{eq:mu_score}
\end{align}

It should be noted that Equation \ref{eq:vaeloss} is identical to the standard VAE loss function. By comparing results using Equations \ref{eq:vaeloss} and \ref{eq:mse_score} with those using Equations \ref{eq:KL_score}, \ref{eq:sigma_score}, and \ref{eq:mu_score}, we directly measure the difference between the full VAE performance compared to the performance of chopping the VAE, which has the added benefit of compressing our network by half.

\paragraph{AD with knowledge ``distillation''}

In order to best evaluate how effective chopping is as a compression technique while retaining performance, we will both (1) compare chopping to KD and (2) combine chopping with KD. We denote scores learned by a student model with $\hat{D}$. Thus, $\hat{D}_\mathrm{MSE}$ and $\hat{D}_\mathrm{MSE+KL}$ denote students regressed to scores from the full model, whereas $\hat{D}_\mu$, $\hat{D}_\sigma$ and $\hat{D}_\mathrm{KL}$ denote students regressed to scores from the chopped model.

For our purposes, we will use a simple and hardware-motivated implementation of knowledge distillation:\marginpar{\emph{distill}} similarly to CICADA \cite{Gandrakota:2024hbo}, we train student models to directly regress the anomaly score produced by the teacher VAE, using the same event-level input variables as the VAE itself. We do not explore the more involved KD techniques discussed in Section \ref{subsec:KD} for two main reasons. First, our teacher architectures are already quite small as they are the same models that will be chopped to fit within the FPGA and therefore our regression task should be feasible without use of more involved KD techniques. Second, because we are making a comparison between chopping and KD, we want to explore techniques that are of comparable effort in terms of developing and deploying a model. We will see that even this simple implementation of KD is sufficient to retain performance in our study. However, the reader should keep in mind that other KD techniques exist that could potentially lead to better performance than demonstrated in this study.

Concretely, for each of the five anomaly score targets defined in Section 3.1.1, we will train several separate student architectures of varying sizes including NN-based architectures and BDT models. The student is trained on the same background events used to train the teacher VAE, minimizing the mean-squared-error loss between the student output and the teacher anomaly score evaluated on those same events. At inference time, the student takes the original event-level inputs and produces a single anomaly score directly, without requiring any part of the VAE. Therefore, if the student architecture is smaller than that of the teacher and can still retain most of the performance of the teacher, we will have successfully distilled and compressed our VAE into a smaller model.

We highlight three important differences from previous work. First, unlike CICADA, we do not perform outlier exposure and thus our students are trained only on the background data and do not see any sort of signal. This is done intentionally to minimize bias of the model towards any specific physics signals. Secondly, we explore KD on all the anomaly scores that we defined in the previous section. Therefore, not only will we compare chopping with the more traditional KD technique where we regress to scores derived from the full VAE teacher, but we will also compare to KD on the chopped scores derived from the latent space. In a way, we utilize the idea that motivates chopping---that only the latent space information is necessary to achieve good AD---within KD and examine if that information can also be transferred effectively to the student models.  Lastly, our student architectures are varied compared to the teacher architecture. Therefore, rather than only creating student architectures that are smaller versions of the teacher architecture, as is commonly done, we also explore different types of architectures, including DNN-based students for CNN VAE teachers and BDT student architectures for both the DNN and CNN VAEs teachers.

\subsection{Datasets}

Datasets in HEP span a wide range of sizes and modalities. Existing uses of AD in triggers have considered both inputs consisting of reconstructed object momenta as well as inputs consisting of energies from a grid of calorimeter cells. In order to study the different methods discussed above we select two datasets that are meant to mimic the different types of input typically used in AD triggers. This choice also allows us to explore the effectiveness of our compression techniques on both CNN and DNN VAE architectures.

\paragraph{Jet Images dataset}

The Jet Images dataset \cite{pierini_2020_3602254} consists of image representations of five classes of jets: quarks, gluons, W bosons, Z bosons, and top quarks. Because the dataset is image-based, the chosen VAE architecture is a CNN VAE. Each image covers a region of $\Delta \eta = \Delta \phi = 2R = 0.16$, centered on the jet axis. This region is binned into 100$\times$100 pixels where each pixel contains the scalar sum of the $p_T$ of the particles in that region, with the highest 150 $p_T$ constituents included for each jet. The angular binning of the dataset is meant to match the typical LHC electromagnetic calorimeter cell sizes

\begin{figure}[ht]
\centering
\includegraphics[width=1\linewidth]{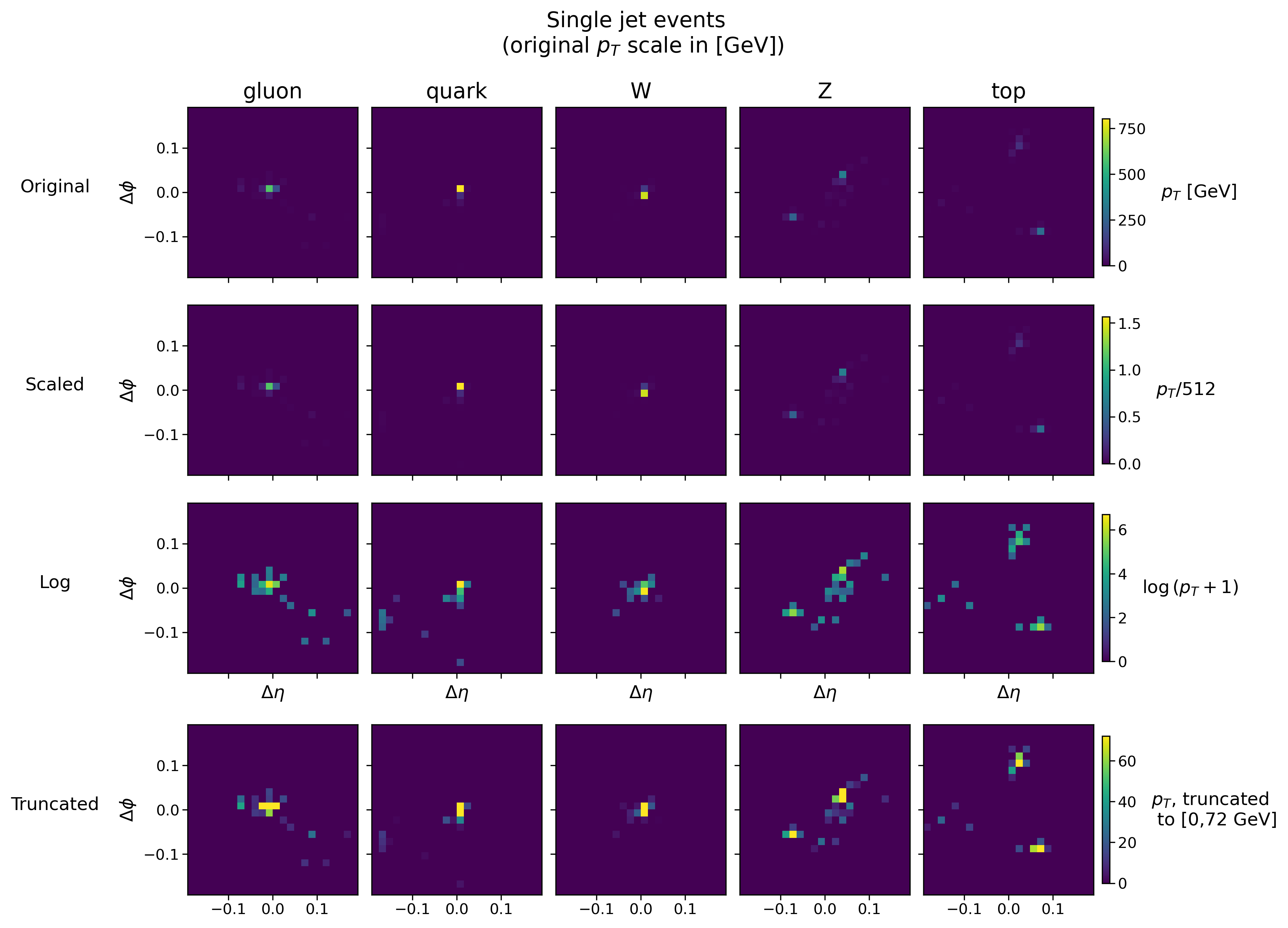}
\caption{Example jet images events with the three different preprocessing techniques: \emph{scaled} where the pixel data is divided by a normalization factor of $512$; \emph{log} where the pixel data is transformed as $\log{(1+p_T)}$; and \emph{truncated} where we assign $p_T > 72$ GeV to that value.
\label{fig:jetimages_preprocessing_techniques}
}
\end{figure}

The images were then further preprocessed. First, the images were cropped to the center 24$\times$24 pixels to avoid overly sparse images while preserving $\gtrsim 95\%$ of the $p_T$ content. Then, since CNN-based models often depend on the preprocessing performed on the image, three techniques, and thus three sets of models, were explored with different purposes in mind (Figure \ref{fig:jetimages_preprocessing_techniques}):
\begin{itemize}
\item \emph{log}: each pixel's $p_T$ is transformed using $\log(1+p_T)$ to compress the distribution range such that lower momentum pixels become more important in training
\item \emph{truncated}: each pixel's $p_T$ is truncated to the 99.5$^\textrm{th}$ percentile of the training data $p_T$ distribution to avoid overly large $p_T$ pixels that might adversely affect model training. The $p_T$ value above the truncation limit is assigned to the largest allowed value.
\item \emph{scaled}: each pixel's $p_T$ is scaled to a smaller value by a factor of 512 to achieve pseudo-normalization in an FPGA-friendly way
\end{itemize}

With this dataset, the goal is to train on the quark and gluon classes and detect W bosons, Z bosons, and top quarks as our anomalies. The images were utilized to train CNN VAEs using an approximate $80\%$ train, $10\%$ test, and $10\%$ validation split on the quark and gluon images \footnote{Only the train .tar file from the dataset was used and this was split into the train, test, and validation sets for this study.}. To determine performance, the three signal classes were combined into a combined signal. A range of 14 VAE architectures were tested for each type of image preprocessing technique, varying the number of layers in the encoder and decoder, the convolutional filter depth, and the dimensions of the latent space. The VAEs were trained to minimize the full VAE loss term (Equation \ref{eq:vaeloss}). The $\beta$ hyperparameter is tuned separately for each preprocessing technique to balance the MSE and KL terms. Values of $\beta=10^{-6}$ for \emph{scaled}, $\beta=10^{-4}$ for \emph{truncated}, and $\beta=10^{-2}$ for \emph{log} were chosen. In addition, a learning rate scheduler and early stopping were utilized.

The full VAE and chopped VAE performance were then evaluated using the five different anomaly scores described by Equations \ref{eq:vaeloss}--\ref{eq:mu_score}.
 
\begin{figure}[ht]
\centering
\includegraphics[width=1\linewidth]{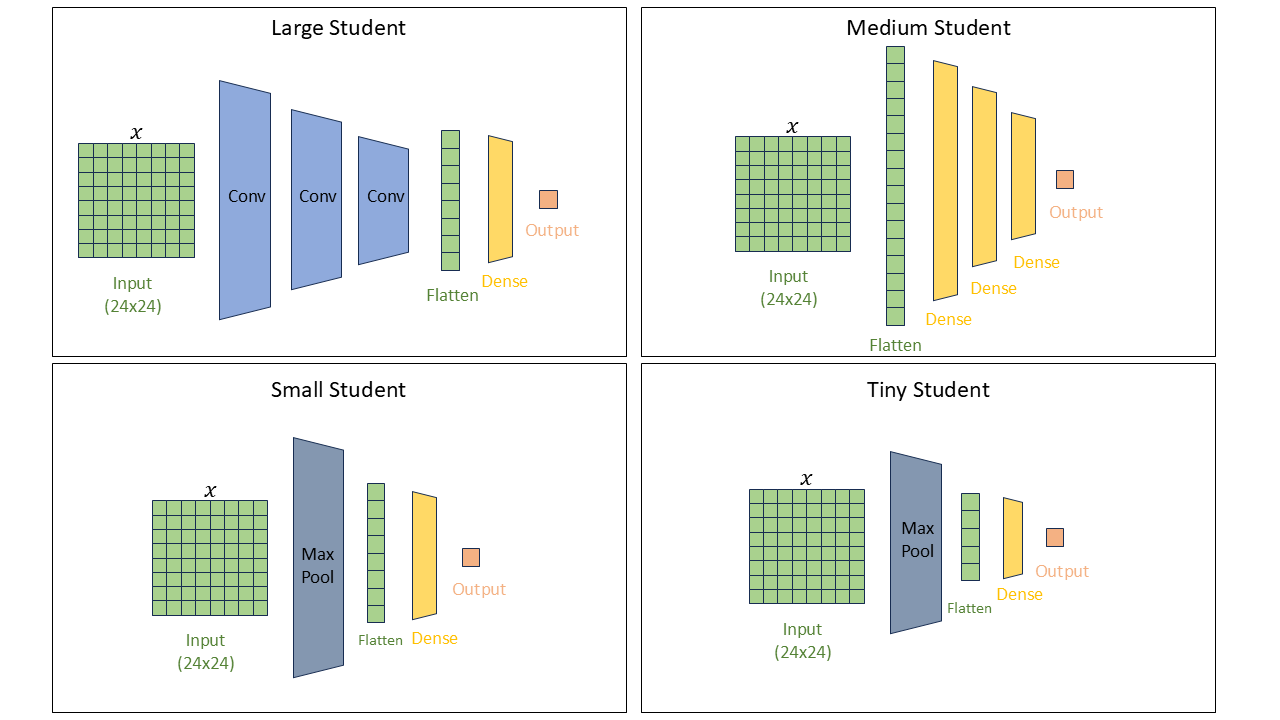}
\caption{
Architecture schematics of the NN student models utilized for the JetImages dataset. The \emph{large} student mimics the encoder structure of the full VAE. The \emph{medium} student flattens the images and passes them through a set of dense layers. The \emph{small} student and the \emph{tiny} student both have max pooling layers of different sizes into a flatten layer and then a dense layer.
\label{fig:jetimages_students_architectures}
}
\end{figure}

Next, knowledge distillation was performed with five types of student architectures; see Figure \ref{fig:jetimages_students_architectures}:

\begin{itemize}
\item \emph{NN-large}: the encoder of the teacher model with more depth
\item \emph{NN-medium}: a flatten layer into three dense layers
\item \emph{NN-small}: a 2$\times$2 max pooling layer into a flatten layer into a dense layer
\item \emph{NN-tiny}: a 3$\times$3 max pooling layer into a flatten layer into a dense layer 
\item \emph{BDT}: a boosted decision tree
\end{itemize}

The general training and evaluation of the student models is similar regardless of if the student is an NN student or a BDT student. The NN student inputs are the full 24x24 preprocessed images used to train the VAE, whereas the BDT student inputs are the same preprocessed images flattened into a 576-dimensional vector. For the NN students, an MSE loss is minimized between the one-dimensional student output and the teacher anomaly score. For the BDT students, TMVA gradient-boosted regression is used, which also minimizes squared error between the predicted and target scores, making the two training objectives equivalent in spirit.

For the BDT students specifically, training is performed using the TMVA~\cite{TMVA:2007ngy} framework with gradient boosting. The BDT configuration uses 200 trees with a maximum depth of 4, a shrinkage (learning rate) of 0.05, stochastic bagging with a subsample fraction of 0.5, and 25 cut points per variable, providing a practical balance between regression fidelity and model complexity.

For each preprocessing technique of \emph{scaled}, \emph{log}, and \emph{truncated}, one well-performing VAE architecture is chosen as the teacher based on strong AD performance in the low FPR region. We then iterate over the five possible anomaly scores and train separate student models to regress to each score. Therefore, we train separate student models for each combination of possible teachers (three options; one for each preprocessing technique), anomaly score targets (five options), and student architectures (five options).

Once the student models are trained, the distillation fidelity and performance are evaluated by computing the anomaly score distributions and ROC curves using the validation background sample and the three signal class samples – top quark, W boson, and Z boson. Comparisons are then made directly to the teacher VAE performance with each anomaly score.

Finally, the models converted to firmware. All of the student models were synthesized with PTQ applied for deployment onto an Alveo U250 FPGA with part number xcu250-figd2104-2L-e. The NN student models were synthesized with HLS4ML using Vivado 2023.2 and a grid search was performed over tunable parameters in HLS4ML, such as reuse factors and implementation strategy, to determine the optimal synthesis settings to generate the resource and latency numbers. The BDT students were converted to firmware using the \texttt{fwXmachina} toolkit; see \emph{Code Availability}, which tabularizes the decision paths of the ensemble into a hardware-executable format 
suitable for deployment on FPGAs at nanosecond-scale latency, as 
discussed further in Section~\ref{sec:method}.

\paragraph{40 MHz dataset}

The 40 MHz dataset \cite{Govorkova:2021syw} was created to study unsupervised detection methods in trigger-like settings. The dataset consists of a selection of common, SM events as a background and includes four different BSM signals that one can attempt to detect. Each event contains $p_T$ (transverse momentum) as well as $\eta$ and $\phi$ (angles specifying the outgoing direction) values for 10 jets, 4 muons and 4 electrons. In addition, the magnitude and $\phi$ of the MET (missing transverse energy) of the event is also given. A lepton filter was applied to all datasets, background and signal, such that each event contains at least one electron, muon, or tau.

We trained a DNN VAE on the $64\%$ of the standard model events, using $16\%$ for validation; the remaining $20\%$ was then used as a background sample when calculating signal efficiencies. The VAE was trained using the full VAE loss function, equation \ref{eq:vaeloss}, with both the MSE and KL Divergence terms. A $\beta$ value close to 0.4 was found to balance the two terms and was thus used. Given the scope of this paper as well as the nature of the 40 MHz dataset, hyperparameters were tuned to optimize for performance in the high-rejection (low False-Positive-Rate) region. 

As always with VAEs, there is a choice to make about the anomaly score; one can use the full loss or just the MSE, which would require a full forward pass through the teacher network. Alternatively, one could use only the KL divergence terms, which requires only half of a forward pass and thus chops the network in half.

We then train student networks on the same standard model events to directly regress to the teacher anomaly scores. We again compare three sizes of neural network students: small (one dense layer), medium (two dense layers), and large (three dense layers). It should be noted that, despite the naming conventions, all three of these students have very small parameter counts; the medium student is roughly the same size as the chopped teacher. We found that the teacher using KL divergence as the anomaly score yielded the highest signal efficiencies in the low False-Positive-Rate region. Accordingly, the KL divergence was used as the target for the neural network student regression; this is contrasted against our more comprehensive JetImages studies. In addition to the neural network students, we also trained a BDT student.

A BDT student is trained for each of the  three anomaly score targets used in this dataset: $\hat{D}_\textrm{MSE}$, $\hat{D}_\textrm{KL}$,  and $\hat{D}_\mu$. As with the  JetImages case, the BDT takes the same event-level kinematic  variables as input to the teacher VAE, namely the transverse momenta,  pseudorapidities, and azimuthal angles of the leading jets, muons,  and electrons, together with the magnitude and azimuthal angle of  the missing transverse energy. No additional feature engineering is  applied beyond what is used for the VAE itself. For the final 40\,MHz BDT study presented here, we use a TMVA gradient boosting configuration with 200 trees, maximum depth of 5, shrinkage of 0.05,  stochastic bagging with a subsample fraction of 0.5, 40 cut  points per variable, and a minimum node size of 0.5\%. To improve the regression fidelity in the high-score tail relevant for the low-FPR trigger regime, events in the upper tail of the target distribution are upweighted during training. The trained BDT students are evaluated on the same four BSM signal samples used for the teacher and neural network student comparisons: $A \to 4\ell$, $H^{\pm} \to \tau\nu$, leptoquark, and $h \to \tau\tau$, alongside the standard model background validation sample, enabling a direct and consistent performance comparison across all student types and all anomaly score targets.

\section{Results}
\label{sec:result}

Results for Jet Images and 40 MHz Datasets are discussed.

\paragraph{Jet Images dataset}

\begin{figure}[ht]
\centering
\includegraphics[width=1.0\linewidth]{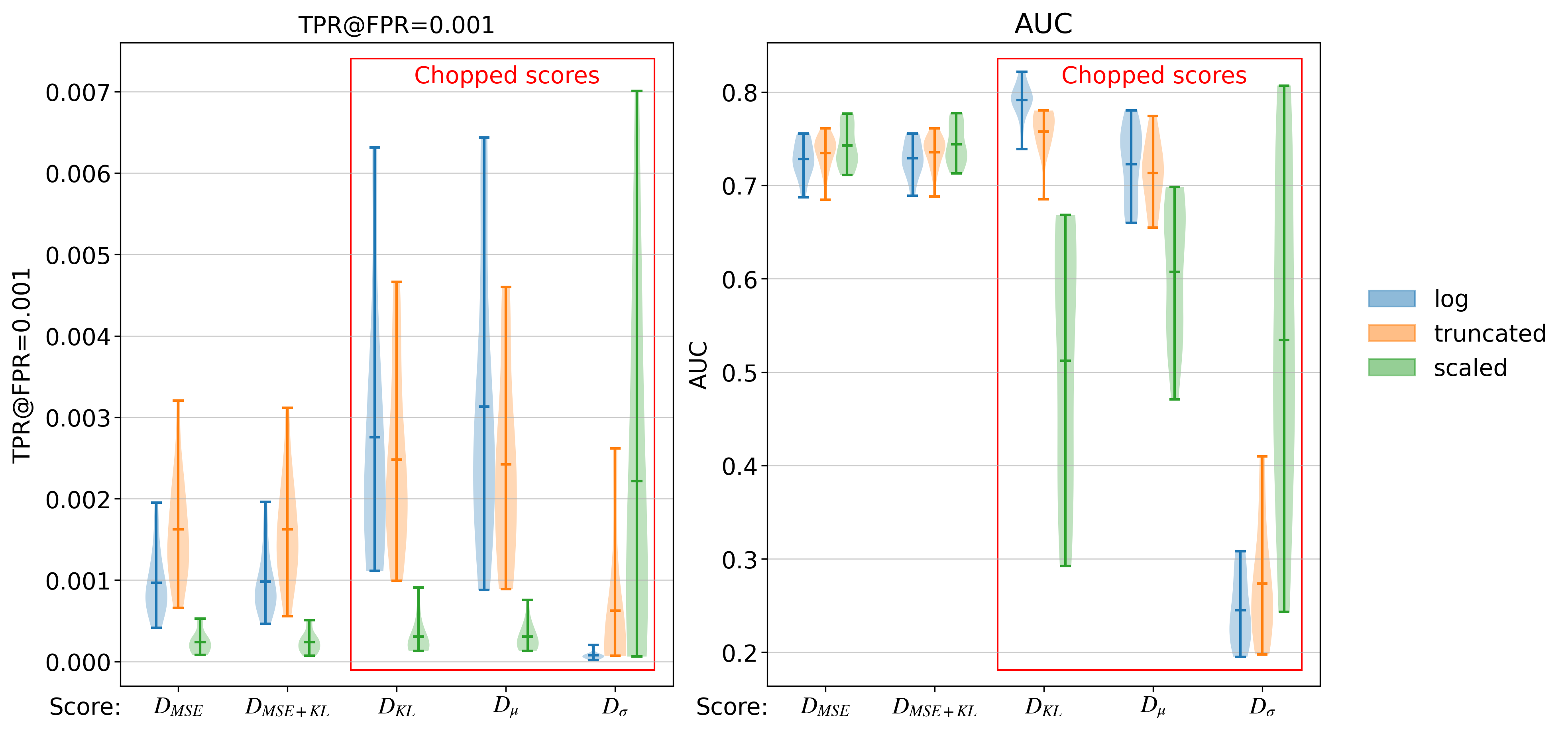}
\caption{
Performance comparison of full model and chopped model AD scores for different VAE architectures. Each violin bar is the summary statistics of fourteen different VAE architectures trained on the jet images with the preprocessing technique given by the color label. The metrics are calculated with the guark and gluon classes as background and with the top, W boson and Z boson classes as the combined signal. The left plot shows the true positive rate at a chosen false positive rate of 0.001. The right plot shows the area under the curve for the full ROC.
\label{fig:jetimages_vae_summary}
}
\end{figure}

First, we discuss the results for chopping. We note that all performance results are shown for the floating-point models and that minimal performance loss can be assumed, as we will discuss, for the final synthesized model. Figure \ref{fig:jetimages_vae_summary} shows the variation in performance of chopping the fourteen different trained VAE models (with different numbers of layers, convolutional filter depth, and latent space dimension) as a function of the different anomaly scores, where $D_\mathrm{MSE}$ and $D_\mathrm{MSE+KL}$ are derived from the full VAE model and $D_\mathrm{KL}$, $D_\mu$, and $D_\sigma$ are derived from the latent space and are therefore the chopped scores. The left plot shows the true positive rate at a chosen false positive rate of 0.001 and the right plot shows the AUC. Across the variety of architectures, we see that chopping can, in many cases, greatly improve performance in the low FPR region that we are generally interested in for trigger applications. However, for many models, this gain can come at the cost of an overall loss in performance as measured by the AUC score.

Further, we find greater variability in both metrics for the chopped scores as compared to the full scores. We also notice that which chopped scores perform well depends significantly on the preprocessing technique used, with $D_\mathrm{KL}$ and $D_\mu$ performing well on \emph{log} and \emph{truncated} models, and $D_\sigma$ performing well on \emph{scaled} models. Therefore, one should take care to scan across architectures, scores, and preprocessing techniques to determine the best models and specific chopped score for each use-case. We conclude that for AD trigger applications, where we are generally interested in performance in the low FPR region, chopping is a very effective method of reducing the model size while maintaining, or in many cases, improving the performance compared to the full VAE architecture.

\begin{figure}[ht]
\centering
\includegraphics[width=\textwidth]{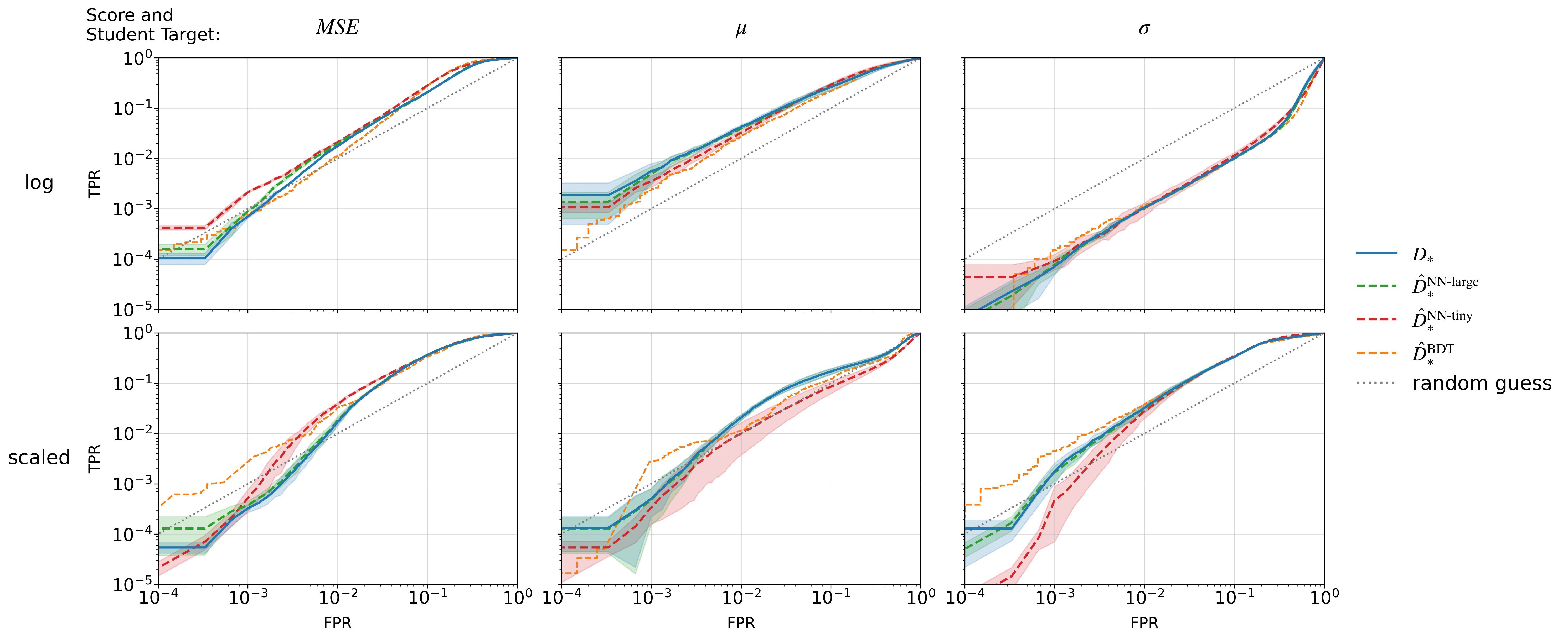}\\
\caption{
ROC plots comparing the performance of three sets of teachers against their students represented by the three columns. The top row shows models trained on \emph{log} transformed inputs and the bottom on \emph{scaled} transformed inputs. Left column: The teacher is the full VAE model with an AD score of $D_\textrm{MSE}$ (solid curve); its students (dashed) are  $\hat{D}_\textrm{MSE}^\textrm{NN-large}$, $\hat{D}_\textrm{MSE}^\textrm{NN-tiny}$, and $\hat{D}_\textrm{MSE}^\textrm{BDT}$. Middle column: The teacher is the chopped model with AD scores of $D_\mu$ (solid curve); its students (dashed) are  $\hat{D}_\mu^\textrm{NN-large}$, $\hat{D}_\mu^\textrm{NN-tiny}$, and $\hat{D}_\mu^\textrm{BDT}$. Right column: The teacher is the chopped model with AD scores of $D_\sigma$ (solid curve); its students (dashed) are  $\hat{D}_\sigma^\textrm{NN-large}$, $\hat{D}_\sigma^\textrm{NN-tiny}$, and $\hat{D}_\sigma^\textrm{BDT}$. Select AD scores are shown to demonstrate the range of results.
\label{fig:jetimages_roc_summary}
}
\end{figure}

Next, we turn to KD. As discussed, KD was performed not only with the full VAE scores of $D_\mathrm{MSE}$ and $D_\mathrm{MSE+KL}$, but also on each of the chopped anomaly scores of $D_\mathrm{KL}$, $D_\mu$, and $D_\sigma$. Figure \ref{fig:jetimages_roc_summary} shows performance results on the combined signals of W, Z, and top for teachers and their students on three of the anomaly score targets (columns) and for two of the preprocessing techniques utilized (rows). We omit $D_\mathrm{MSE+KL}$ and $D_\mu$ because these models perform similarly to $D_\mathrm{MSE}$ and $D_\mu$, respectively, and the medium and small students as they perform similarly to large and tiny students, respectively. We find that, regardless of architecture and preprocessing technique, most of the student architectures generally mimic the teacher’s performance fairly well. The large (and medium) student follow the teacher the closest, whereas the rest of the students either match the teacher's performance or perform slightly better or worse, depending on the specific architecture and the FPR region. In some cases, we find that students can even outperform the teacher, which is most notable with the tiny and BDT students. Lastly, as noted in Figure \ref{fig:jetimages_vae_summary}, we again see that the performance with each score depends on the preprocessing technique utilized. In particular, $D_\sigma$ performs very poorly for the \emph{log} models, but very well for the \emph{scaled} models. Overall, we find that KD is a very effective method of further reducing model size while maintaining, or in some cases, improving upon the performance of the teacher. We find that some of the best-performing models are those that combine the chopped anomaly scores with KD.

\begin{figure}[ht]
\centering
\includegraphics[width=1.0\textwidth]{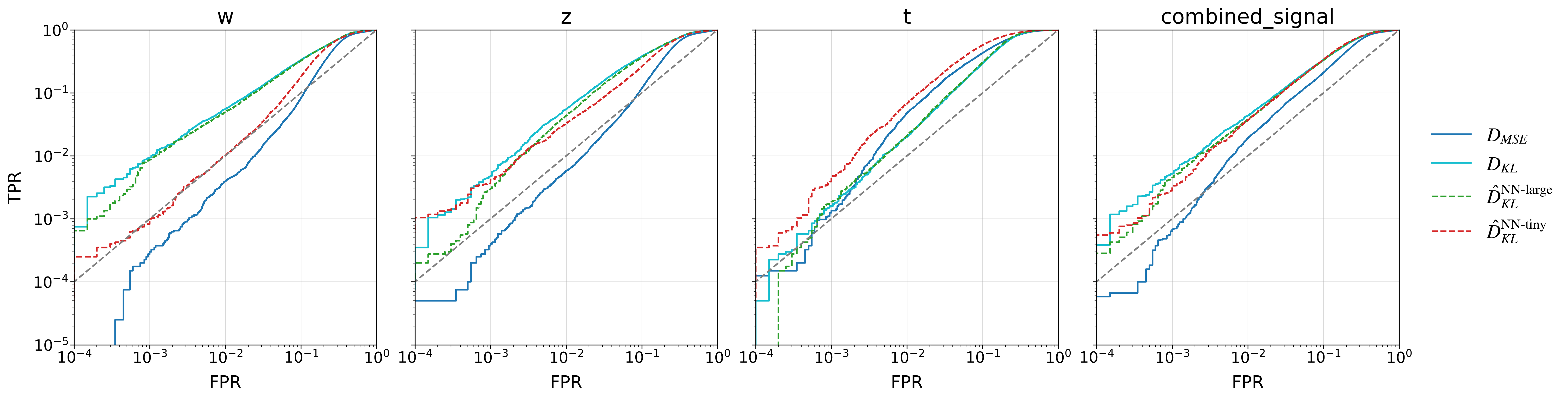}\\
\caption{
Comparison of the performance of the full VAE model with score $D_\textrm{MSE}$ to the chopped model with score $D_\textrm{KL}$ and two of its student models: $\hat{D}_\textrm{KL}^\textrm{NN-large}$ and $\hat{D}_\textrm{KL}^\textrm{NN-tiny}$. The models shown are trained on the \emph{log} transformed input data.
\label{fig:jetimages_student_variability_plot}
}
\end{figure}

However, we find some variability in the student performance. Whereas Figure \ref{fig:jetimages_roc_summary} only showed the performance for the combined signals, Figure \ref{fig:jetimages_student_variability_plot} shows the performance for the three signals separately and compares the full VAE ($D_\mathrm{MSE}$) with the chopped model ($D_\mathrm{KL}$), and the distilled chopped students ($\hat{D}_\mathrm{KL}$). We find that the chopped model and the large student perform relatively similarly on the different signals of W, Z, and top, whereas the tiny student is variable, performing much worse on W, slightly worse on Z, and better on top compared to the chopped teacher model. We omit the medium and small students in this figure, but note that they perform similary to the large student (see the full Figure \ref{app:jetimages_signal_variability}). This per-signal variability is not visible in the combined signal curves, but demonstrates a point of caution in reducing the student size too much. Even a small difference in architecture size, such as that between the small and tiny student, can result in a large difference in per-signal performance. This general idea could very well be an example of the capacity gap between teacher and student described in \cite{mirzadehImprovedKnowledgeDistillation2020} and could therefore potentially be addressed with intermediate distillation steps. This is especially important in the use-case of anomaly detection as we generally have no labels of anomalies, and so these per-signal effects can be entirely hidden while looking at performance metrics. Therefore, erring on the side of caution while reducing KD student size is recommended, as well as utilizing techniques such as intermediate distillation steps.

\begin{figure}[ph!]
\centering
\vspace{-0.1in}\includegraphics[width=0.95\textwidth]{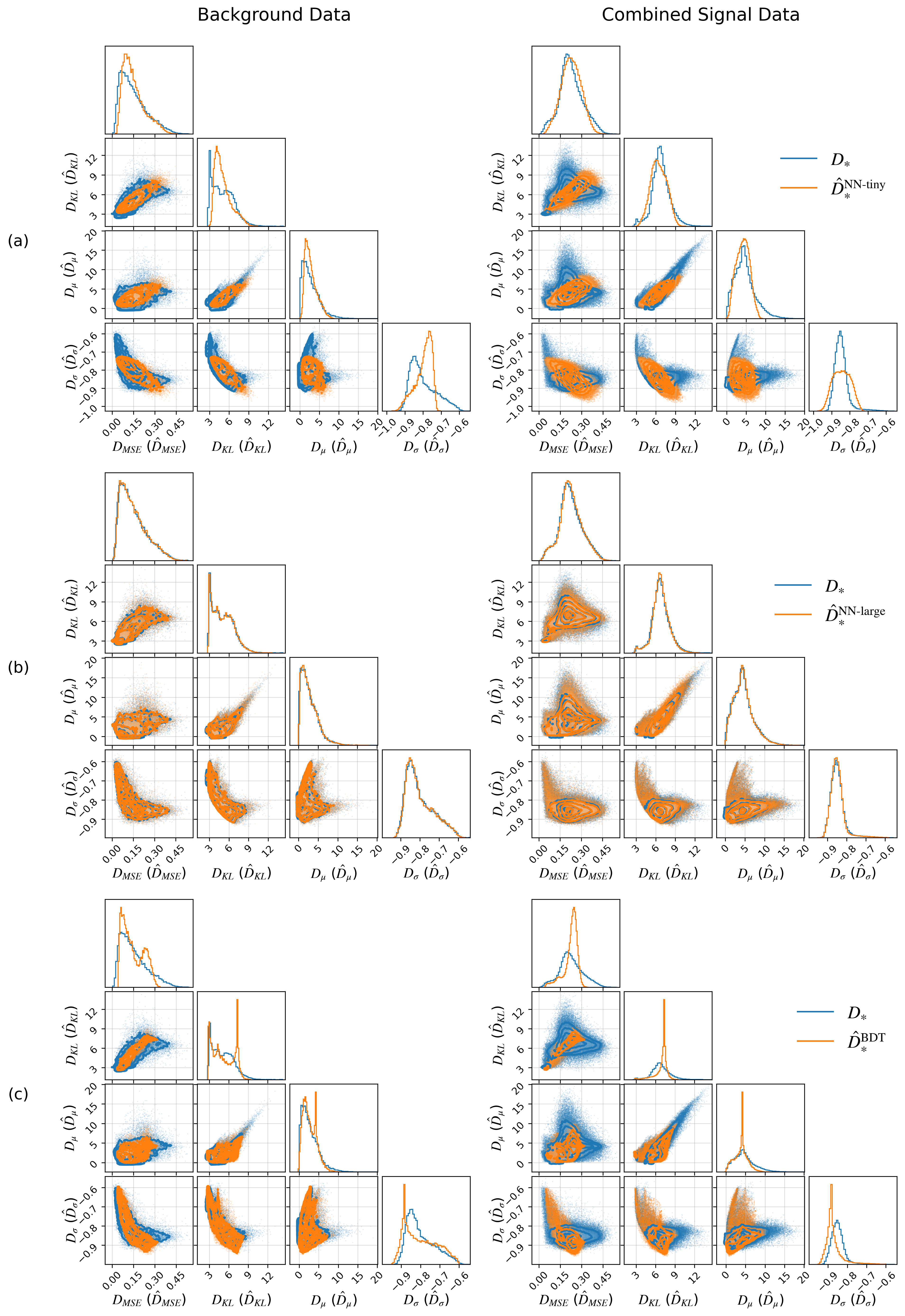}\\
\caption{
Comparison of pairwise AD scores (pairings of MSE, KL, $\mu$, $\sigma$) from models trained on \emph{log} input data are shown as 2d contours; 1d distributions are also shown. The six sets are from the student types from three rows and two columns: (a) NN-tiny, (b) NN-large, (c) BDT; and data for background (left column) and combined signal (right column). Each pairing shows blue contours for the teacher and orange contours for the students. Each axis denotes the teacher (student) AD score as $D$ ($\hat{D}$). NB. The teacher contours for the three rows (a,b,c) are identical whereas the student contours are unique to that row.
\label{fig:jetimages_contours_combined}
}
\end{figure}

To further explore the nuances in student performance, we plotted 2d pairwise distributions of different anomaly score targets for the teacher (in blue) and the student (in orange) in Figure \ref{fig:jetimages_contours_combined}. The left column shows the background data that the models were trained on, whereas the right column shows the combined signal data. The different rows compare the (a) NN-tiny, (b) NN-large, and (c) BDT student models to the teacher models. First, by comparing the left column with the right column, we can qualitatively see the separation between background and signals in the score space. We then find that as the student size decreases from large to tiny, there is a shift in student behavior: when the student model is small enough, there is not enough capacity to perfectly learn the teacher scores. Instead, the student model will learn the general trends and even 2D correlations will remain between different student models regressed to different scores. This trend is also clearly seen in the BDT student. These contours can perhaps explain why small student models can sometimes outperform the teacher: the small student models smooth out the teacher distributions by learning more general patterns, rather than more specific information. However, as noted in the previous paragraph, one must reduce the student size cautiously to avoid introducing additional per-signal variability in the small student models.A notable artifact specific to the BDT students on the jet images input is the quantization of the predicted score distribution. Whereas NN students produce smooth, continuous output distributions, the BDT output collapses to a small number of discrete values, visible as sharp spikes in the one-dimensional score histograms. This arises from an inductive bias mismatch: the VAE computes its anomaly scores through many layers of nonlinear convolution, so the scores are a highly nonlinear function of the raw pixel values. Events that differ widely in their true anomaly score can therefore appear nearly identical in pixel space, causing the BDT to assign them to the same leaf and the same predicted value. We find that increasing the number of trees from 200 to 500 and reducing the learning rate do not resolve this artifact (Pearson $r$ changes by less than 0.003), consistent with the bottleneck being in the mismatch between the BDT inductive bias and the nonlinear structure of the CNN-derived scores rather than in model capacity. Crucially, however, quantized scores do not degrade anomaly detection performance: ROC curves and AUC values remain essentially identical between the BDT student and the teacher, because anomaly detection depends on the \emph{ranking} of events, not on the absolute score values. The discrete outputs also map naturally to lookup tables in firmware, which is an advantage for FPGA deployment. This behavior contrasts with the 40\,MHz dataset, where the BDT student inputs are already high-level kinematic features rather than raw pixels,
and no such quantization artifact appears.

\begin{figure}[ht]
\centering
\includegraphics[width=0.95\textwidth]{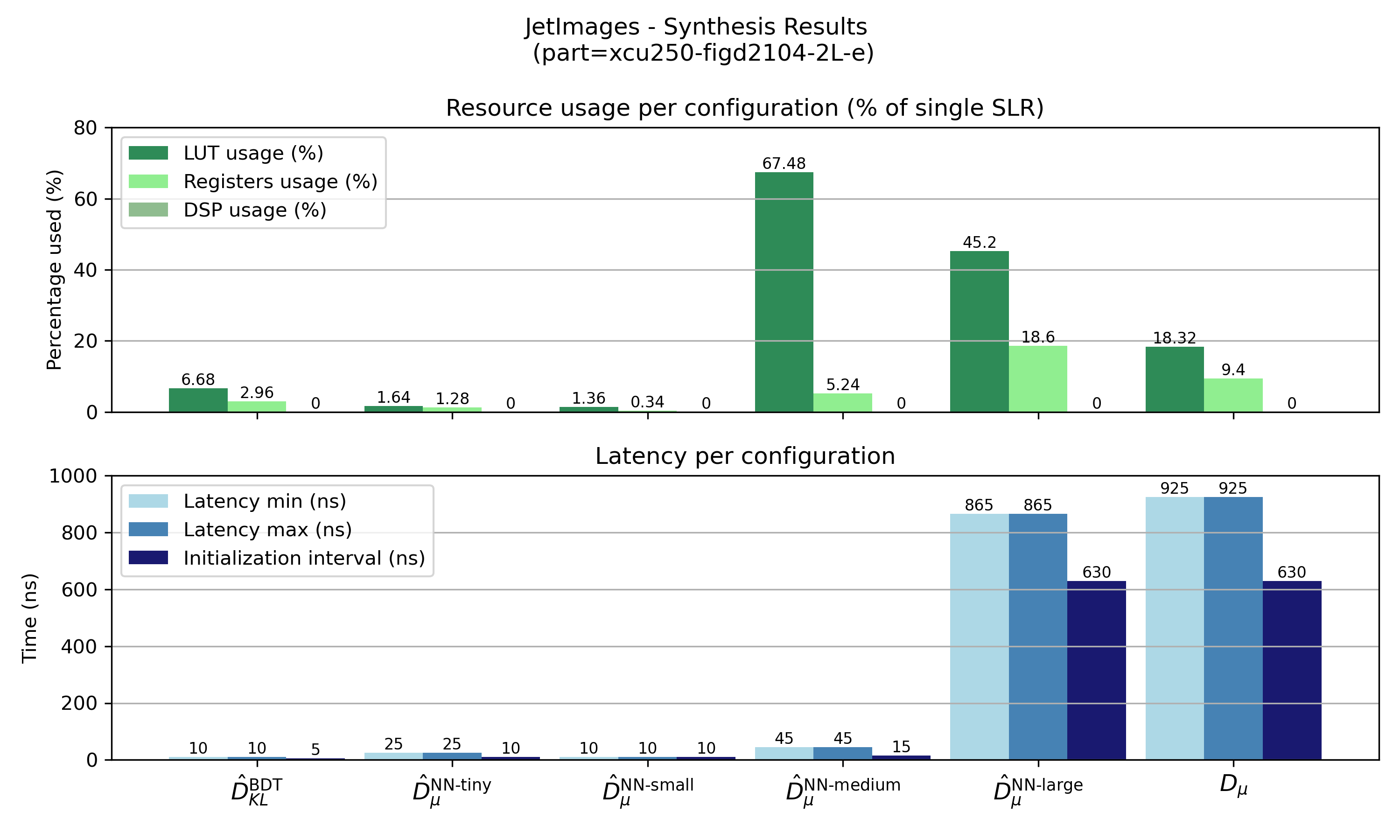}\\
\caption{
The resource (per SLR) and latency estimates for the chopped model and student models of the Jet Images dataset. All the models are synthesized with post training quantization for deployment on an Alveo U250 FPGA with part number xcu250-figd2104-2L-e. The NN estimates are made with HLS4ML using Vivado 2023.2. The BDT estimates are made with fwXmachina and are implemented directly in Vivado 2020.2 as generated VHDL. Note that the NN-medium student resource usage is relatively high and can be reduced to $< 10\%$ with quantization-aware training (QAT) methods. In general, these results show an upper-bound in resource and latency usage, which can be further reduced through QAT.
\label{fig:jetimages_synthesis_results}
}
\end{figure}

\paragraph{40 MHz dataset}

\begin{figure}[htbp]
\centering
\includegraphics[width=\textwidth]{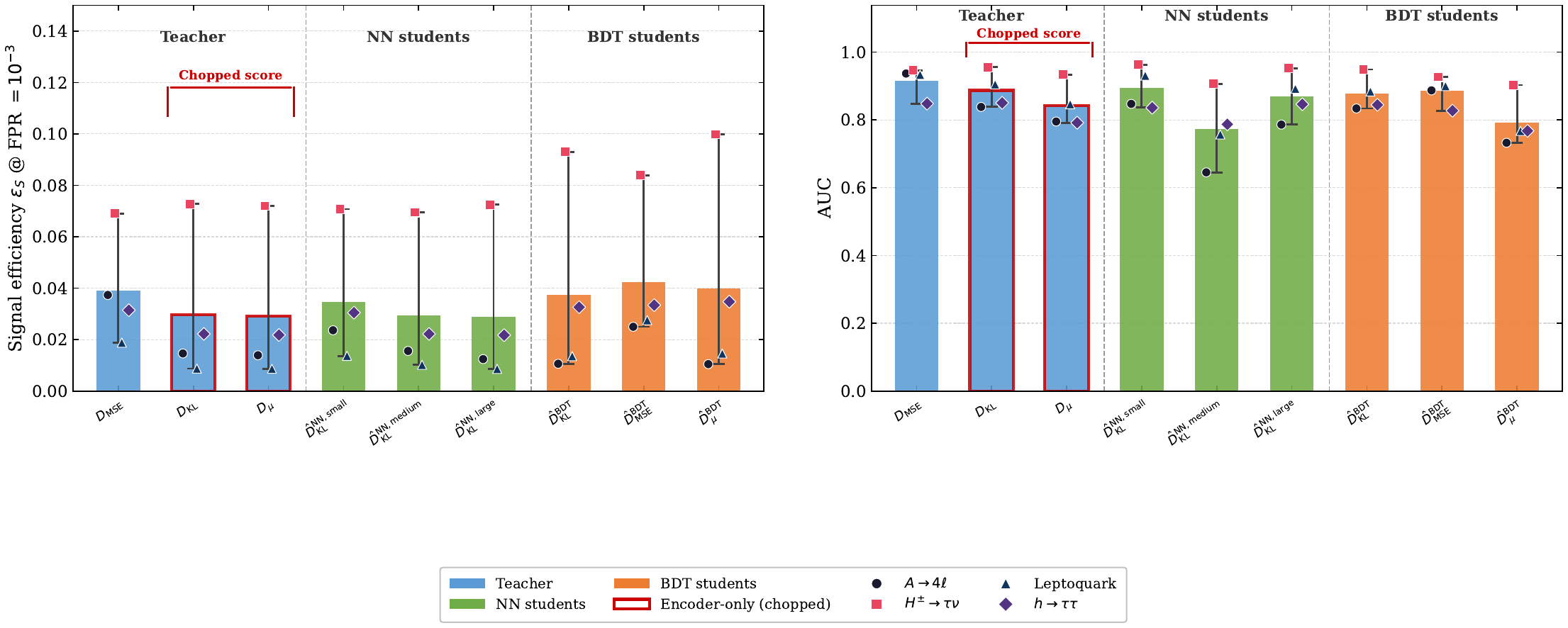}
\caption{
Performance summary for all compression methods on the 40\,MHz dataset across the four BSM signal benchmarks. Left: signal efficiency $\varepsilon_S$ at the trigger working point $\mathrm{FPR} = 10^{-3}$. Right: area under the ROC curve (AUC). For each method, the filled bar height is the \emph{mean} over the four BSM signals; grey error bars span the minimum and maximum across those signals; and each colored marker shows the value for one benchmark ($A\to 4\ell$, $H^{\pm}\to\tau\nu$, leptoquark, $h\to\tau\tau$). Bar color denotes model class: blue (VAE teacher), green (NN student), orange (BDT student). The red outline and bracket mark the \emph{chopped} encoder-only teacher scores (KL and $\mu$), which require only the VAE encoder at inference; MSE alone requires the decoder. The full VAE (MSE) attains the highest working-point efficiency on most signals ($\varepsilon_S$ up to 3.7\% for $A\to 4\ell$). Among compressed models, the small NN student and the BDT (MSE target) reach the highest $\varepsilon_S$ on $A\to 4\ell$ ($2.4\%$ and $2.5\%$), while the BDT (MSE target) closes the AUC gap to the teacher most effectively (AUC\,=\,0.887 vs.\ 0.937). Signal-to-signal spread here reflects a single trained model evaluated on four BSM benchmarks; this differs from the JetImages summary (Fig.~\ref{fig:jetimages_vae_summary}), where the envelope reflects variability across independently trained architectures.
\label{fig:40mhz_summary}
}
\end{figure}

Figure~\ref{fig:40mhz_summary} gives an overview of the performance of all compression methods across the four BSM signal benchmarks, showing both the signal efficiency at the trigger working point chosen for this dataset ($\mathrm{FPR} = 10^{-3}$) and the full-ROC AUC. As with the JetImages dataset, all performance results shown below are for the floating-point models; we discuss the impact of quantization and synthesis at the end of this section.

\begin{figure}[htbp]
\centering
\subfloat[Note that chopping reduces the AUC with respect to the MSE.]{\includegraphics[width=0.75\textwidth]{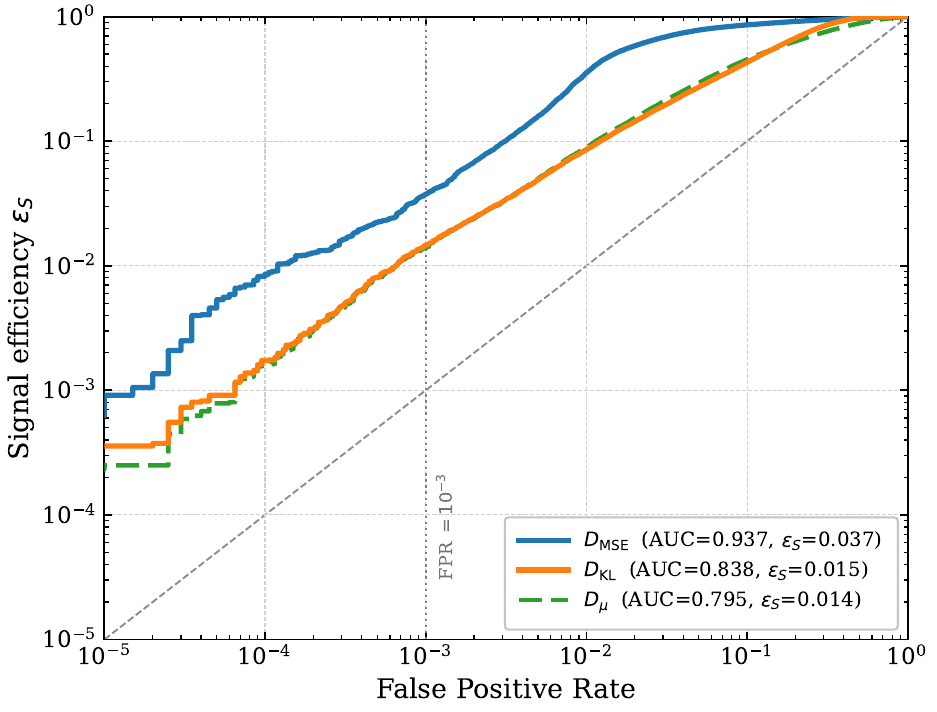}}
\hfill
\subfloat[Note that the small NN student outperforms the teacher.]{\includegraphics[width=0.75\textwidth]{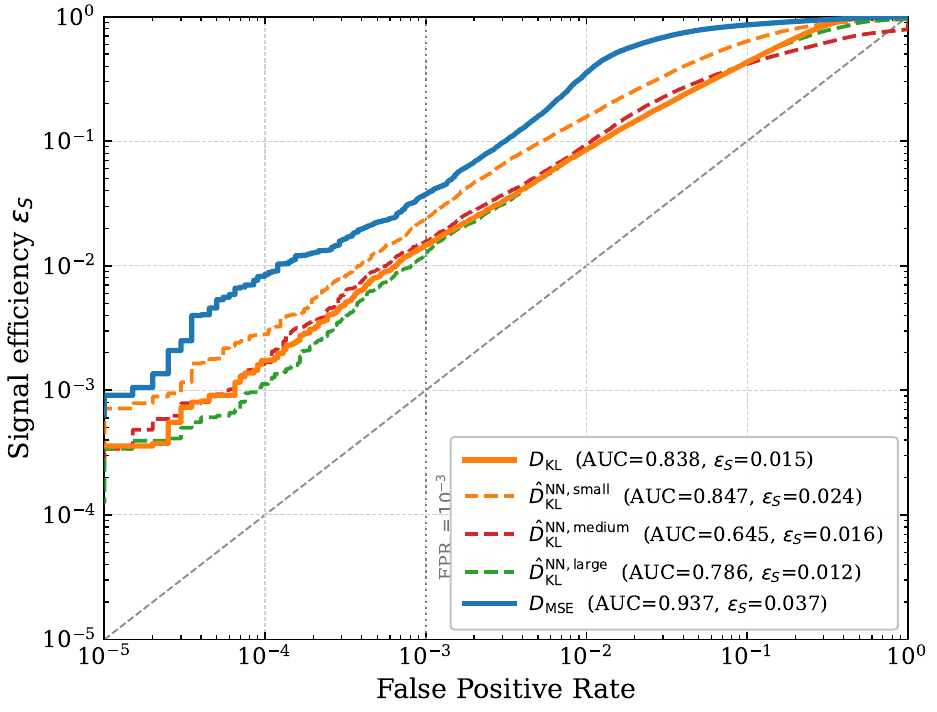}}
\caption{
ROC curves for $A\to 4\ell$ signal of the 40\,MHz dataset. (a) Full VAE ($D_\textrm{MSE}$) vs.\ the two encoder-only chopped scores ($D_\textrm{KL}$ and $D_\mu$). (b) $D_\textrm{KL}$ teacher score vs.\ $\hat{D}_\textrm{KL}^\textrm{NN}$ students trained to regress it (small, medium, large), with the MSE score shown as an upper-bound reference. The $D_\textrm{MSE}$ and $D_\textrm{KL}$ in solid curves in both panels are identical. The vertical dotted line indicates FPR$=10^{-3}$.
\label{fig:40mhz_chop_kd}
}
\end{figure}

We first discuss the results for chopping. For the 40\,MHz dataset, only the MSE reconstruction loss requires the VAE decoder at inference time; both the KL divergence and the encoder-only $\mu$ score are computed entirely from the encoder outputs ($\mu_i$, $\sigma_i$) and therefore represent the chopped model. Figure~\ref{fig:40mhz_chop_kd}(a) compares these three anomaly scores, quantifying the cost of removing the decoder. We find that chopping incurs a moderate penalty: the signal efficiency at the working point drops from 3.7\% (MSE) to 1.5\% (KL) and 1.4\% ($\mu$) for the $A \to 4\ell$ benchmark, while the AUC falls from 0.937 to 0.838 and 0.795 respectively. The small gap between KL and $\mu$ in the low-FPR regime ($\varepsilon = 1.5\%$ vs.\ 1.4\%) indicates that the choice of encoder-only score has little impact on trigger-relevant performance. The much larger gap relative to MSE instead reflects the genuine information cost of discarding the decoder reconstruction. It should be noted that, among all four signals, $A \to 4\ell$ had the largest drop in performance from chopping. We therefore conclude, as for the JetImages case, that chopping is a viable strategy for reducing the model to half its original size, at the expense of a small to moderate performance penalty that depends on the signal topology.

We next turn to knowledge distillation. As previously mentioned, all three NN student architectures are trained to regress the KL anomaly score (the chopped teacher target). Figure~\ref{fig:40mhz_chop_kd}(b) shows the KL teacher score alongside the three NN student architectures for the $A \to 4\ell$ signal; the MSE baseline is included for reference. In general, the students learn to approximate the KL teacher reasonably well, though with notable architecture-dependent variation. The small student (AUC\,=\,0.847, $\varepsilon\,{=}\,2.4\%$) outperforms both the large student (AUC\,=\,0.786, $\varepsilon\,{=}\,1.2\%$) and the medium student (AUC\,=\,0.645, $\varepsilon\,{=}\,1.6\%$) at the working point. The inversion between small and large students mirrors the behaviour described in the Jet Images results: when the student capacity is insufficient to replicate the teacher score precisely, the model instead learns smoother, more general discriminants that can be advantageous in the low-FPR regime. The medium student performs significantly worse than either of the other students across all four signals (AUC\,=\,0.645 for $A \to 4\ell$), suggesting an unfavorable capacity for this particular architecture on the 40\,MHz event representation. As with the JetImages results, this per-signal variability is largely hidden when looking at aggregate metrics, and we therefore caution against reducing the student size too aggressively without checking individual signal performance.

\begin{figure}[htbp]
\centering
\subfloat[Note that at FPR$=10^{-3}$, the two student ($D_\textrm{MSE}^\textrm{BDT}$, $D_\textrm{KL}^\textrm{NN-small}$) perform comparably.]{\includegraphics[width=0.75\textwidth]{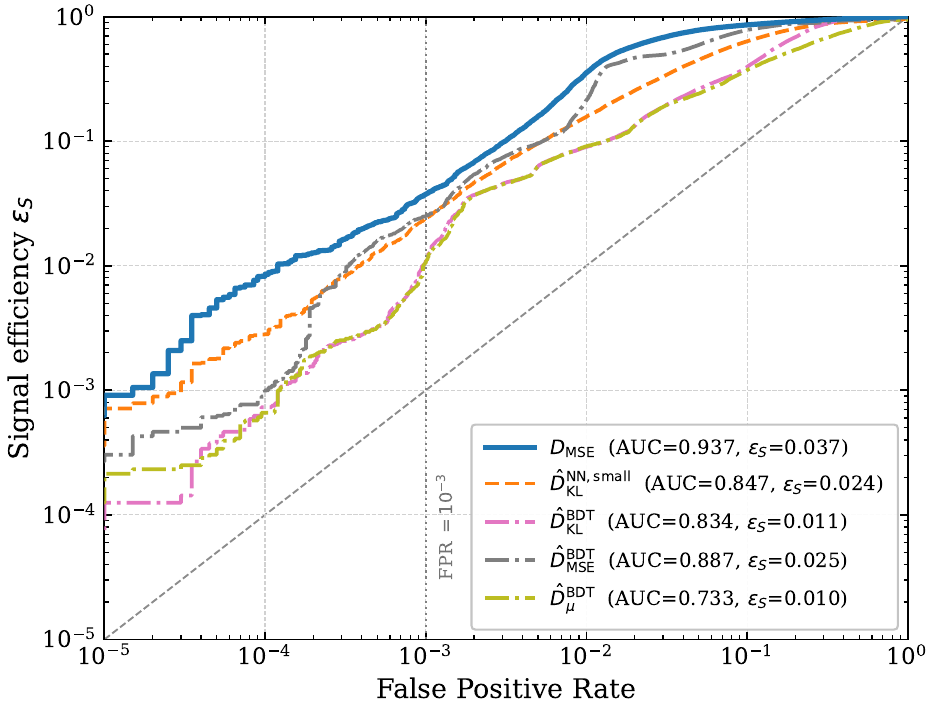}}
\hfill
\subfloat[Note that at FPR$=10^{-3}$, the two students ($\hat{D}_\textrm{KL}^\textrm{NN-small}$, $\hat{D}_\textrm{MSE}^\textrm{BDT}$) are the highest performing. ]{\includegraphics[width=0.75\textwidth]{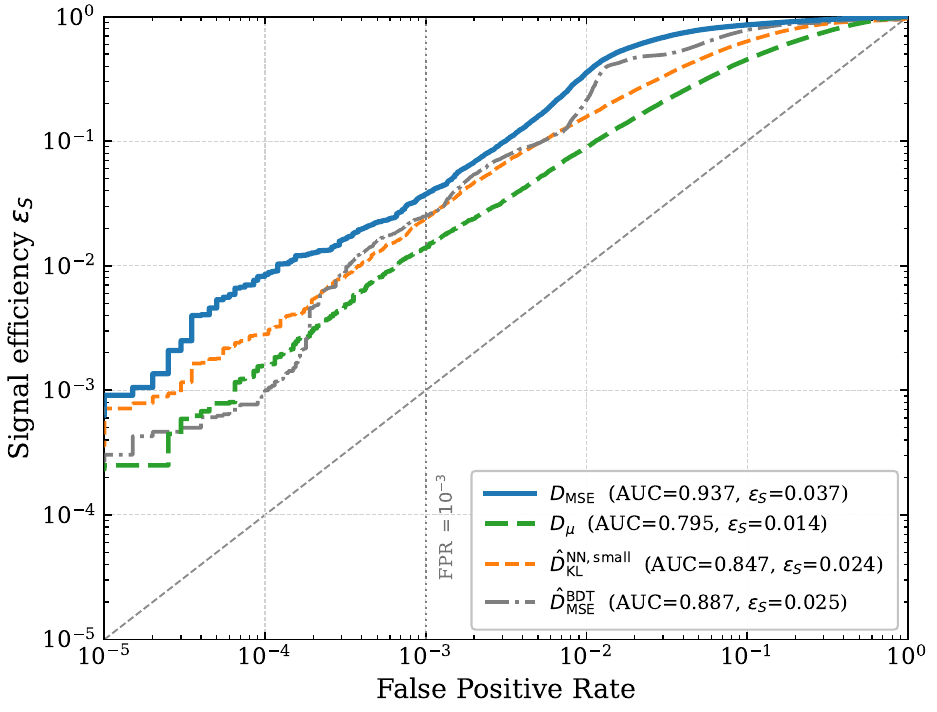}}
\caption{
ROC curves for the $A\to 4\ell$ signal of the 40\,MHz dataset. (a) Full VAE ($D_\textrm{MSE}$) vs.\ four students (three BDT and one NN). (b) Head-to-head comparison of one representative from each compression class: full VAE ($D_\textrm{MSE}$), chopped encoder-only ($D_\mu$), small NN student ($\hat{D}_\textrm{KL}^\textrm{NN-small}$), and BDT student ($\hat{D}_\textrm{MSE}^\textrm{BDT}$). The vertical dotted line marks $\mathrm{FPR} = 10^{-3}$.
\label{fig:40mhz_hardware}
}
\end{figure}

Figure~\ref{fig:40mhz_hardware}(a) compares the BDT regression students against the small NN student and the MSE teacher. We find that the choice of regression target has a large impact on BDT performance. The BDT trained on the MSE target achieves the highest AUC among all student models (0.887), approaching within 5\% of the MSE teacher, with working-point efficiency $\varepsilon = 2.5\%$ for $A \to 4\ell$ (comparable to the small NN student at 2.4\%). BDT trained on the KL and $\mu$ targets reach lower AUC on this signal (0.834 and 0.733 respectively), with working-point efficiencies of $\varepsilon = 1.1\%$ and 1.0\% for $A \to 4\ell$. Across the four benchmarks, the BDT students can exceed the corresponding teacher at the working point on individual signals (e.g.\ $\varepsilon = 9.3\%$ vs.\ 7.3\% for the BDT (KL target) on $H^{\pm} \to \tau\nu$), while the MSE-target BDT remains the most balanced compressed model in AUC.

Figure~\ref{fig:40mhz_hardware}(b) collects one representative from each compression class for a direct head-to-head comparison, which most directly addresses the central question of this paper. At the trigger working point, the small NN student retains $\sim$65\% of the full VAE's efficiency ($\varepsilon = 2.4\%$ vs.\ 3.7\%), while the chopped score ($\mu$) remains lower at $\sim$38\% ($1.4\%$). The BDT (MSE target) recovers a comparable fraction ($\varepsilon = 2.5\%$, $\sim$68\%) and is therefore competitive with the small NN student despite the very different architecture.

Table~\ref{tab:40mhz_efficiency} summarises the signal efficiency at $\mathrm{FPR} = 10^{-3}$ across all four BSM benchmarks for every method. The $H^{\pm} \to \tau\nu$ benchmark shows the most uniform performance across methods ($\varepsilon = 6.9$--$10.0\%$), indicating that this signal is well-separated from background regardless of model size, while $A \to 4\ell$ shows the widest spread and is the most sensitive benchmark for distinguishing compression strategies.

\begin{figure}[ht]
\centering
\includegraphics[width=0.95\textwidth]{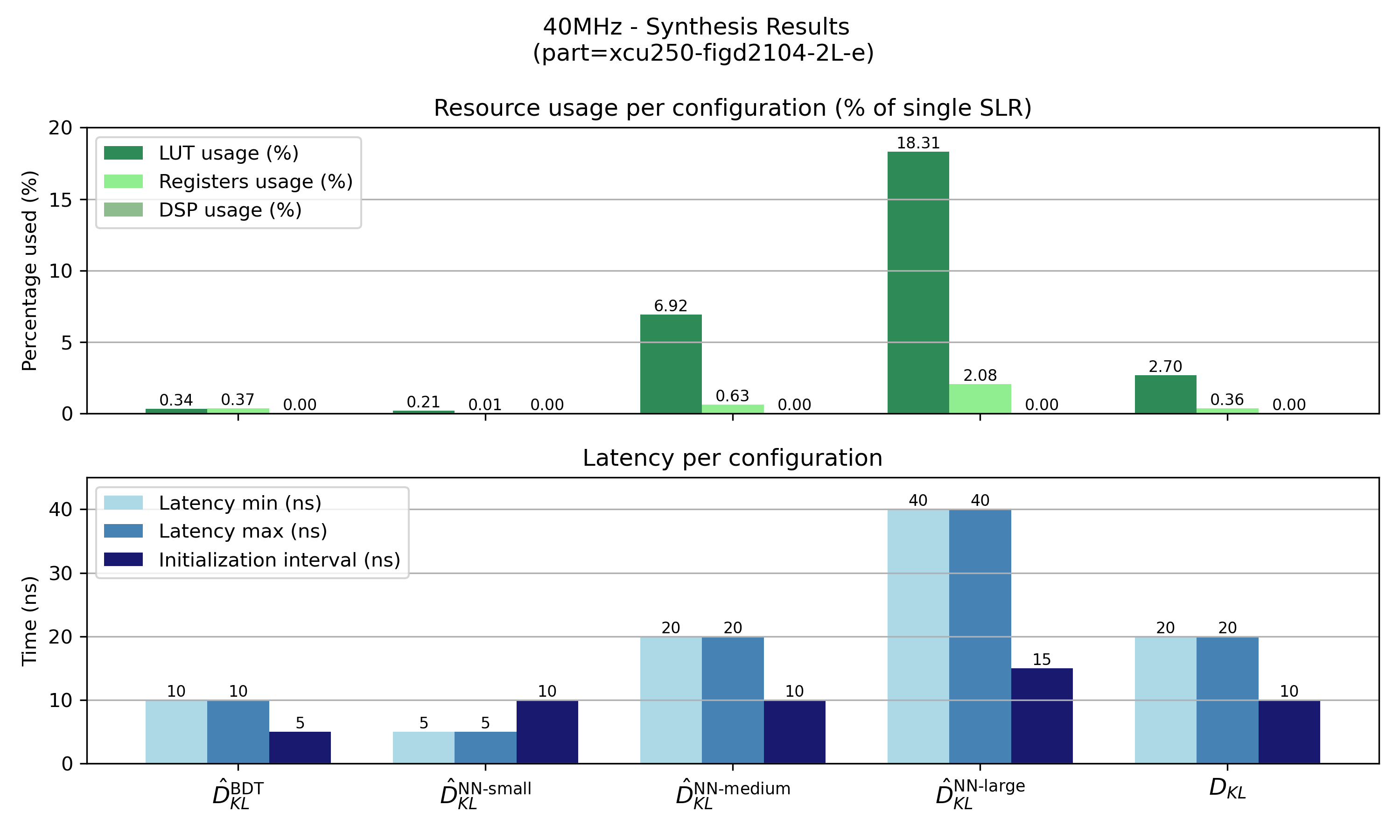}\\
\caption{
The resource (per SLR) and latency estimates for the student models of the 40 MHz dataset. All the models are synthesized with post training quantization for deployment on an Alveo U250 FPGA with part number xcu250-figd2104-2L-e. The NN estimates are made with HLS4ML using Vivado 2023.2. The BDT estimates are made with fwXmachina and are implemented directly in Vivado 2020.2 as generated VHDL. In general, these results show an upper-bound in resource and latency usage, which can be further reduced through QAT.
\label{fig:40Mhz_synthesis_results}
}
\end{figure}

We now discuss the synthesis results, comparing the resource and latency requirements for the student models on the 40\,MHz dataset. Figure \ref{fig:40Mhz_synthesis_results} shows the resource usage and latencies for the different models after PTQ was applied to each. As with the JetImages case, we utilized PTQ as a simple way to estimate the results one might obtain if the models were retrained using more rigorous quantization-aware techniques. We find that all of the student models fit within reasonable LHC FPGA trigger algorithm constraints, though with significant variation across architectures. The large student is by far the most resource intensive, using 18.32\% of the LUTs and 2.04\% of the registers of a single SLR, with a latency of 40\,ns. The medium student is substantially smaller, requiring 6.92\% LUT usage and 0.64\% register usage, with a latency of 20\,ns. In contrast, the small student and the BDT student are both extremely lightweight, each using well below 1\% of the resources of a single SLR. The small student has the lowest latency at 5\,ns, while the BDT student has a latency of 10\,ns and an initialization interval of 5\,ns. We also note that none of the student models require DSP resources in this implementation. Overall, these results show that the compressed student models, and in particular the small NN and BDT students, can satisfy aggressive latency and resource constraints for FPGA deployment on the 40\,MHz task.

\begin{table}[htbp]
\centering
\caption{
Signal efficiency $\varepsilon_S$ (\%) at $\mathrm{FPR} = 10^{-3}$ for all compression methods across the four 40\,MHz BSM signal benchmarks; ratios are given with respect to the Full Teacher in the first row. Both KL and $\mu$ are the ``chopped'' encoder-only scores; only MSE requires the VAE decoder. The $D$ ($\hat{D})$ denotes the anomaly score (distilled score) associated with the model. Statistical uncertainties on $\varepsilon_S$ are at most $0.5\%$ absolute. Horizontal rules separate the three model classes.
\label{tab:40mhz_efficiency}
}
\begin{tabular}{lll c cccc c cccc}
\hline
&&&&
\multicolumn{4}{l}{$\varepsilon_S$ (\%)\dotfill}
&&
\multicolumn{4}{l}{Ratio wrt Full Teacher\dotfill}
\\
Score&Architecture&Role &
& $A_{4\ell}$
& $H^{\pm}_{\tau\nu}$
& LQ 
& $h_{\tau\tau}$
&
& $A_{4\ell}$
& $H^{\pm}_{\tau\nu}$
& LQ 
& $h_{\tau\tau}$ \\
\hline
MSE score\\
\quad $D_\textrm{MSE}$                         & Full              & Teacher && 3.7 & 6.9 & 1.9 & 3.1 && 1   & 1   & 1   & 1\\
\quad $\hat{D}_\textrm{MSE}^\textrm{BDT}$      & Distilled         & Student && 2.5 & 8.4 & 2.8 & 3.3 && 0.7 & 1.2 & 1.5 & 1.1\\
KL score\\
\quad $D_\textrm{KL}$                          & Chopped           & Teacher && 1.5 & 7.3 & 0.9 & 2.2 && 0.4 & 1.1 & 0.5 & 0.7\\
\quad $\hat{D}_\textrm{KL}^\textrm{NN-large }$ & Distilled chopped & Student && 1.2 & 7.2 & 0.9 & 2.2 && 0.3 & 1.0 & 0.5 & 0.7\\
\quad $\hat{D}_\textrm{KL}^\textrm{NN-medium}$ & Distilled chopped & Student && 1.6 & 6.9 & 1.0 & 2.2 && 0.4 & 1.0 & 0.5 & 0.7\\
\quad $\hat{D}_\textrm{KL}^\textrm{NN-small }$ & Distilled chopped & Student && 2.4 & 7.1 & 1.4 & 3.0 && 0.6 & 1.0 & 0.7 & 1.0\\
\quad $\hat{D}_\textrm{KL}^\textrm{BDT}$       & Distilled chopped & Student && 1.1 & 9.3 & 1.4 & 3.3 && 0.3 & 1.3 & 0.7 & 1.1\\
$\mu$ score\\
\quad $D_\mu$                                  & Chopped           & Teacher && 1.4 & 7.2 & 0.9 & 2.2 && 0.4 & 1.0 & 0.5 & 0.7\\
\quad $\hat{D}_\mu^\textrm{BDT}$               & Distilled chopped & Student && 1.0 & 10. & 1.5 & 3.5 && 0.3 & 1.4 & 0.8 & 1.1 \\
\hline
\end{tabular}
\end{table}

\section{Summary and Outlook}
\label{sec:conclusion}

While AD in trigger has only begun to be explored in HEP, two main deployment strategies are already beginning to develop. In this work we have set out to compare these approaches, in an effort to understand whether there is a clear choice that should be selected for future AD trigger deployment. As with many AD studies, we find that there is no single approach that works best for each signal and each type of input. Chopping, distillation, and their combination often perform quite well. We conclude that these approaches which are currently deployed in AD triggers are viable methods of reducing VAE architectures in terms of latency and resources. However, we do find that with some tuning KD is often capable of slightly outperforming chopping alone. The degree of this improvement is again dependent on the signal, and often requires  studies as we have presented to ensure that the student architecture is properly selected. In some cases, KD on the chopped results is even capable of outperforming the teacher and other student models, although this result is not guaranteed. We suggest that the success of KD for AD can be attributed to its ability to smooth the results of the teacher and shift probability mass from tails to the bulk, thereby lowering the overlap between poorly reconstructed signals and background. This provides a mechanism for selecting a model without the use of signals, solely through the failure of a student to exactly reproduce the distribution of its teacher. While KD is potentially capable of producing very performant AD triggers, we note that chopping is ultimately an extremely simple and, in many cases, roughly equivalent method for hardware deployment, and should not be seen as inferior. Predicting the success of a particular method, whether chopping, KD, or both, remains difficult. An improved theoretical understanding of these strategies would help maximize the sensitivity of future AD triggers to potential new physics. While our studies have focused on VAE-based AD, we note that non-VAE architectures can also produce powerful anomaly detectors, and that their deployment could also potentially benefit from KD; the study of these architectures is left for future work.

\backmatter

\bmhead{Acknowledgements}
We thank Yuvaraj Elangovan of the Dietrich School Electronics Shop Core Facility at the University of Pittsburgh for consultation on boosted decision tree firmware.

\section*{Declarations}

\subsection*{Funding}

MC and DR are supported by the U.S. Department of Energy (DOE), Office of Science, Office of High Energy Physics Early Career Research program under Award No.\ DE-SC0025324.  TMH, RG, and SAS are supported by the U.S. Department of Energy Award No.\ DE-SC0007914. This work used resources available through the National Research Platform (NRP) at the University of California, San Diego~\cite{10.1145/3708035.3736060}.  NRP has been developed, and is supported in part, by funding from National Science Foundation, from awards 1730158, 1540112, 1541349, 1826967, 2112167, 2100237, and 2120019, as well as additional funding from community partners. Work performed in Dietrich School Electronics Shop Core Facility (RRID:SCR 025113) and services and instruments used in this project were graciously supported, in part, by the University of Pittsburgh. 

\subsection*{Competing interests}

One author declares competing interests. TMH is part of a patent application on the firmware design of the boosted decision trees with the University of Pittsburgh as US Patent Application Publication No.\ US 2024/0054399. Other authors declare no competing interests.

\subsection*{Data availability}

The datasets used in this research have already been made publicly available and are cited in the text.

\subsection*{Code availability}

The code is available in the following Github repository (\url{https://github.com/rajat116/choppers}).

\subsection*{Author contribution}

MC, RG, and SH performed model training and produced performance results. SH and SS produced FPGA implementation results. SH coordinated the overall effort. TMH and DSR conceived of and supervised the project. All authors contributed to writing manuscript text.

\noindent

\begin{appendices}

\section{Performance Plots}
\label{app:jetimages}
\label{app:40mhz_full}

\begin{figure}[htbp]
\centering
\includegraphics[width=1\textwidth]{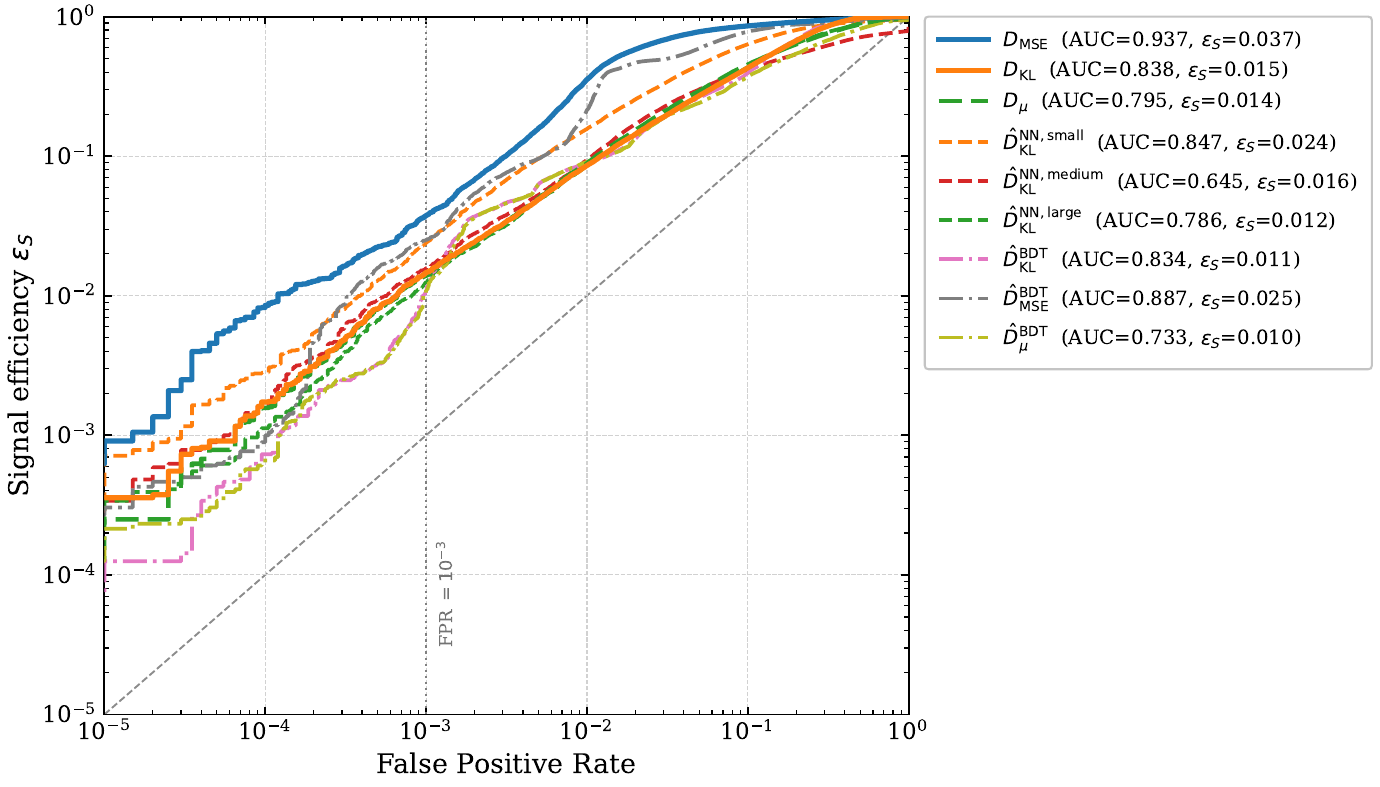}
\caption{
Full ROC comparison for the $A\to 4\ell$ signal. This benchmark shows the widest performance spread across methods (AUC\,=\,0.645--0.937), making it the most diagnostic for comparing compression strategies. The MSE teacher achieves the highest working-point efficiency ($\varepsilon\,{=}\,3.7\%$), while the small NN student ($\varepsilon\,{=}\,2.4\%$) and the BDT (MSE target) ($\varepsilon\,{=}\,2.5\%$) are the strongest compressed models at $\mathrm{FPR}=10^{-3}$. The BDT (MSE target) reaches the highest AUC among students (0.887). The medium student performs significantly worse than all other models (AUC\,=\,0.645, $\varepsilon\,{=}\,1.6\%$), consistent with a capacity gap for this signal topology.
\label{fig:supp_ato4l}
}
\end{figure}

\begin{figure}[htbp]
\centering
\includegraphics[width=1\textwidth]{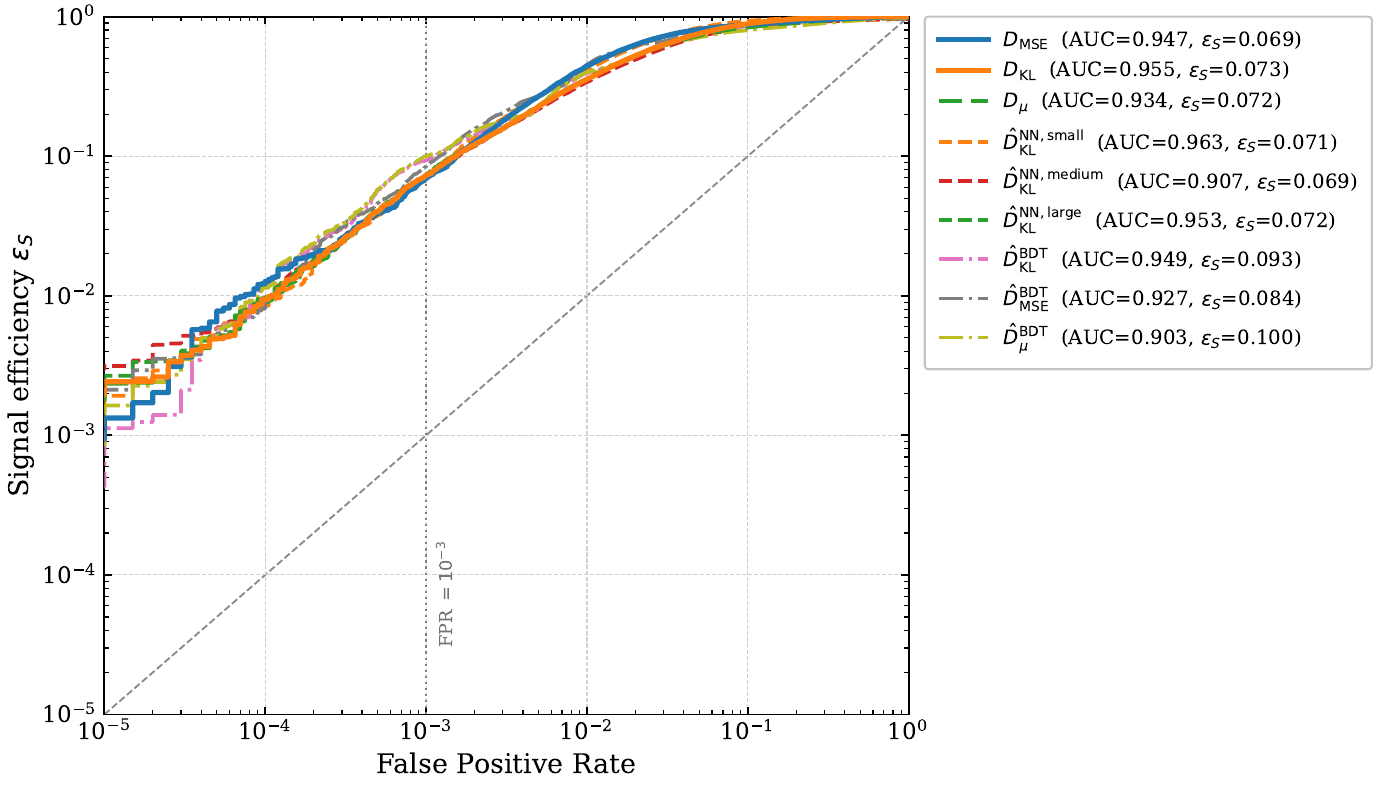}
\caption{
Full ROC comparison for the $H^{\pm}\to\tau\nu$ signal. All methods perform well and are closely clustered (AUC\,=\,0.903--0.963, $\varepsilon\,{=}\,6.9$--$10.0\%$), indicating this signal is well-separated from background regardless of model size or compression strategy. Notably, the small NN student achieves the highest overall AUC (0.963), exceeding the teacher (0.947), consistent with the smoothing effect of limited model capacity discussed in the Jet Images results. The KL teacher achieves the highest working-point efficiency among the teacher and NN models ($\varepsilon\,{=}\,7.3\%$), while the BDT students reach even higher working-point efficiencies, with $\varepsilon\,{=}\,9.3\%$ for the KL target and $\varepsilon\,{=}\,10.0\%$ for the $\mu$ target.
\label{fig:supp_hChToTauNu}
}
\end{figure}

\begin{figure}[htbp]
\centering
\includegraphics[width=1\textwidth]{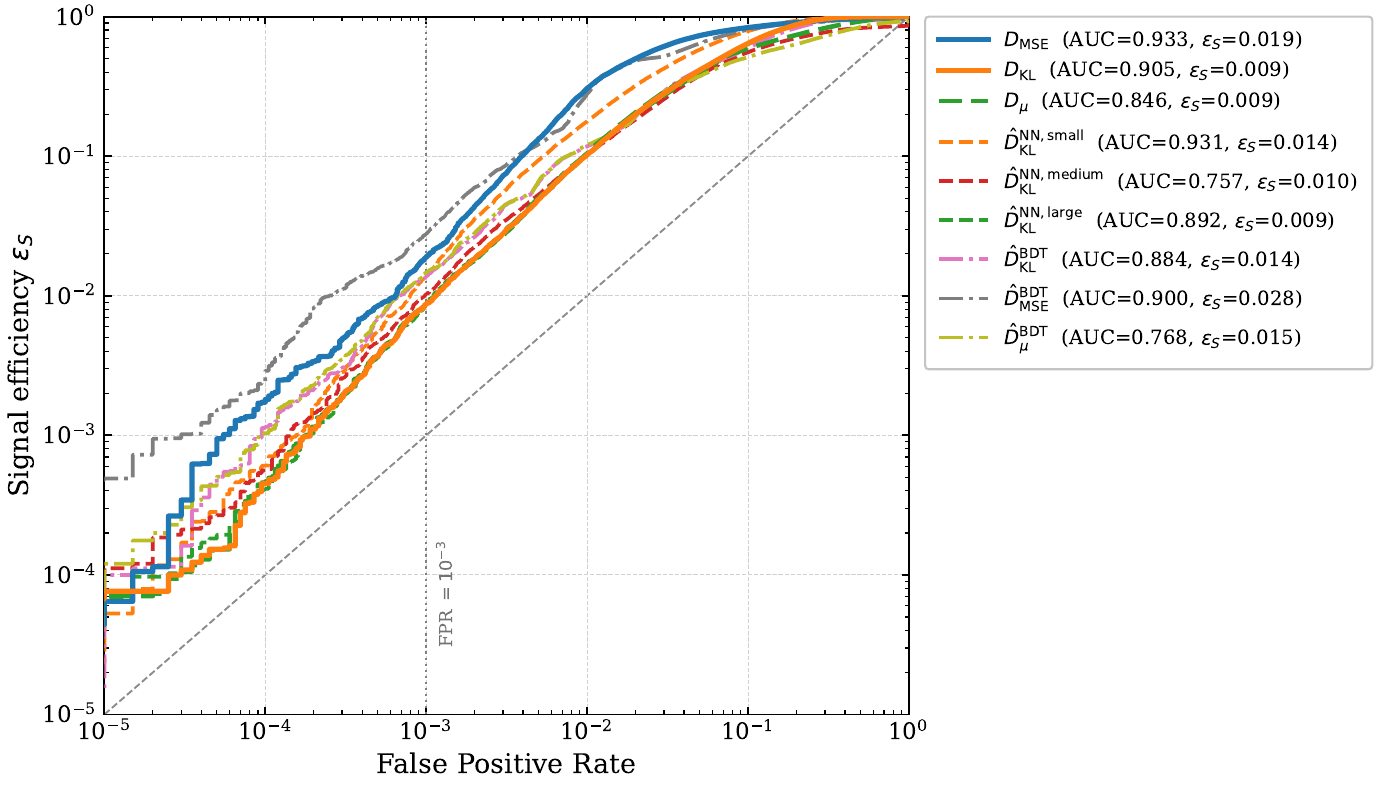}
\caption{
Full ROC comparison for the leptoquark signal. The MSE teacher achieves the highest AUC (0.933). The small NN student (AUC\,=\,0.931, $\varepsilon\,{=}\,1.4\%$) closely matches the teacher in both metrics, while the BDT (MSE target) reaches the highest working-point efficiency ($\varepsilon\,{=}\,2.8\%$) with AUC\,=\,0.900. The medium student again underperforms (AUC\,=\,0.757, $\varepsilon\,{=}\,1.0\%$) relative to the other students, while BDT trained on the KL and $\mu$ targets remain weaker in AUC but competitive at the working point.
\label{fig:supp_leptoquark}
}
\end{figure}

\begin{figure}[htbp]
\centering
\includegraphics[width=1\textwidth]{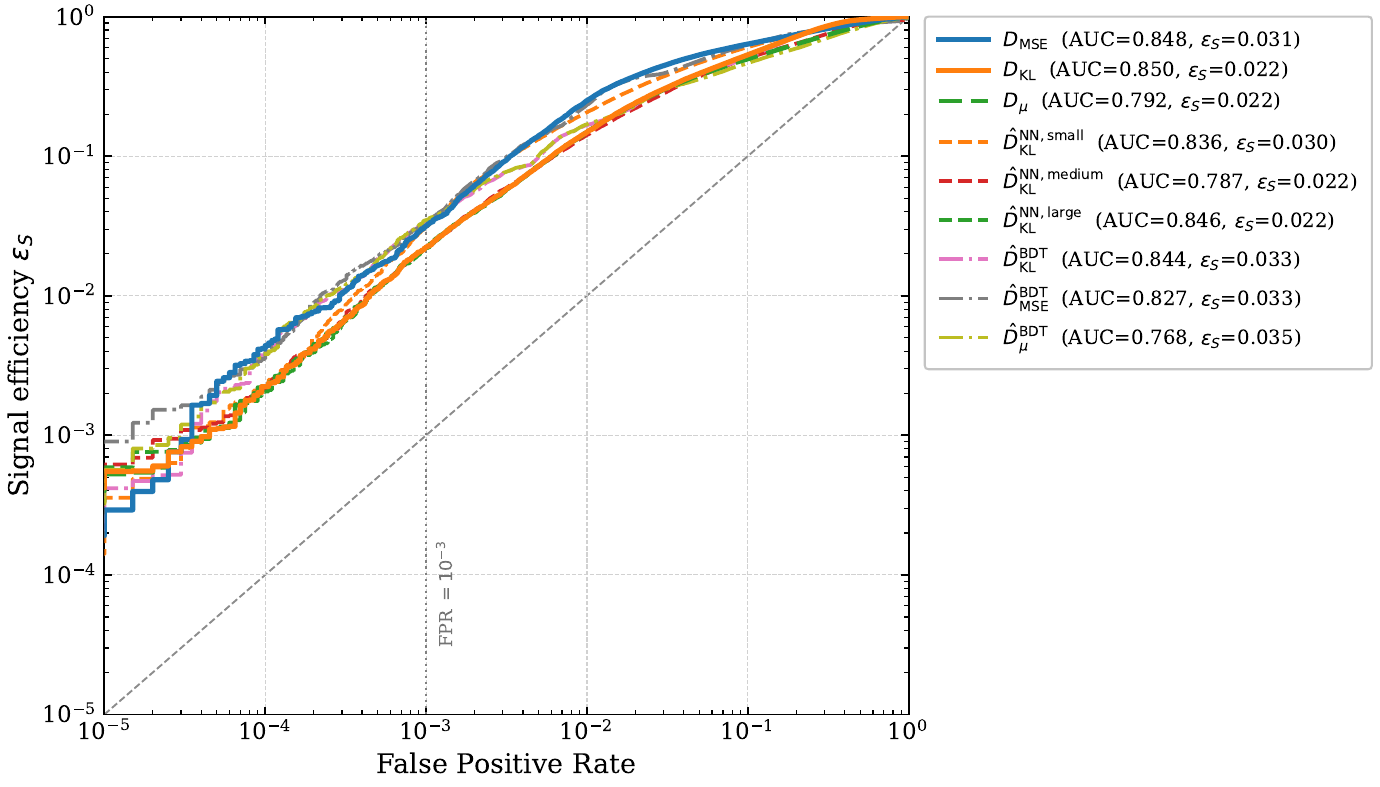}
\caption{
Full ROC comparison for the $h\to\tau\tau$ signal. Overall performance is modest and the methods are closely bunched (AUC\,=\,0.768--0.850), reflecting the difficulty of separating this signal from the Standard Model background. The KL teacher achieves the highest AUC (0.850), while the MSE teacher retains the best working-point efficiency among the teacher models ($\varepsilon\,{=}\,3.1\%$). The small NN student closely tracks the teacher at the working point ($\varepsilon\,{=}\,3.0\%$), while the BDT students reach $\varepsilon\,{=}\,3.3$--$3.5\%$, illustrating again that working-point efficiency and full-ROC AUC need not rank models in the same order.
\label{fig:supp_htautau}
}
\end{figure}

Plots for per-signal variability plot for Jet Images dataset are shown. as well as the ROC comparisons for 40\,MHz dataset. Figures~\ref{fig:supp_ato4l}--\ref{fig:supp_htautau} show ROC curves for all nine compression methods across each of the four 40\,MHz BSM signal benchmarks. AUC values and signal efficiencies at $\mathrm{FPR}=10^{-3}$ are quoted in each legend entry.

\begin{figure}[ht]
\centering
\includegraphics[width=1.0\textwidth]{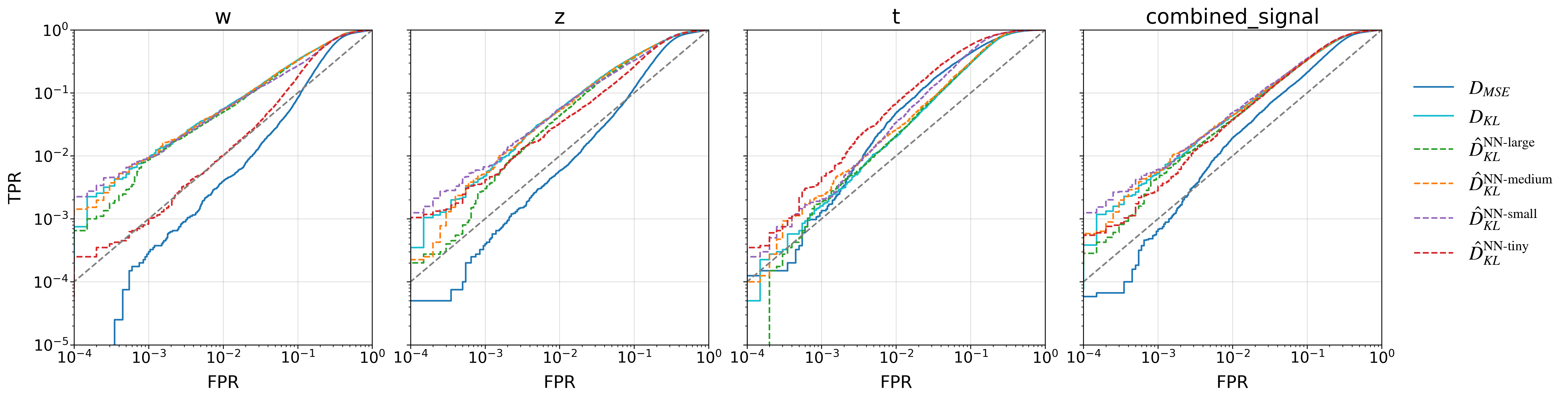}\\
\caption{
Comparison of the performance of the full VAE model with $D_\textrm{MSE}$ as the AD score with the performance of the chopped model with $D_\textrm{KL}$ as the AD score, and all the NN student models, each regressed to KL as the AD score. The models shown are are trained on the \emph{log} transformed input data.
\label{app:jetimages_signal_variability}
}
\end{figure}

\end{appendices}

\FloatBarrier

\bibliography{references}

\end{document}